\documentclass[showpacs, prd, twocolumn, preprintnumbers, nofootinbib]{revtex4}

\usepackage{amsmath}
\usepackage{amssymb}
\usepackage{color}
\usepackage{verbatim}
\usepackage{graphicx}

\newcommand{\dif}{\mathrm{d}}
\newcommand{\bb}{\bar{B}}
\newcommand{\tg}{\tilde{g}}
\newcommand{\te}{\tilde{\eta}}
\newcommand{\tk}{\tilde{\kappa}}
\newcommand{\tb}{\tilde{\beta}}
\newcommand{\hk}{\hat{k}}

\newcommand{\doubarrow}[1]{\overset{\leftrightarrow}{#1}}
\newcommand{\he}{\hat{\eta}}

\begin{document}

\title{Generalized bumblebee models and Lorentz-violating electrodynamics}

\author{Michael D.\ Seifert}

\affiliation{Dept.\ of Physics, Indiana University, 727 E.\
$\text{3}^\text{rd}$ St., Bloomington, IN, 47405}

\email{mdseifer@indiana.edu}

\begin{abstract}
  The breaking of Lorentz symmetry via a dynamical mechanism, with a
  tensor field which takes on a non-zero expectation value in vacuum,
  has been a subject of significant research activity in recent years.
  In certain models of this type, the perturbations of the
  ``Lorentz-violating field'' about this background may be identified
  with known forces.  I present the results of applying this
  interpretation to the ``generalized bumblebee models'' found in a
  prior work.  In this model, the perturbations of a Lorentz-violating
  vector field can be interpreted as a photon field.  However, the
  speed of propagation of this ``bumblebee photon'' is
  direction-dependent and differs from the limiting speed of
  conventional matter, leading to measurable physical effects.  Bounds
  on the parameters of this theory can then be derived from resonator
  experiments, accelerator physics, and cosmic ray observations.
\end{abstract}

\preprint{IUHET 533, September 2009}



\maketitle

\section{Introduction}

The experimental signatures of a violation of Lorentz symmetry have
been extensively sought for in recent years (see \cite{LVdata} and
references therein.)  The primary paradigm for examining the physics
of such effects is the ``Standard Model Extension'' (SME)
\cite{CPTviol, CKSME}.  Broadly speaking, in the usual picture of the
Standard Model, one writes down a list of field combinations that are
renormalizable and invariant under Lorentz symmetry (as well as under
various other desired symmetries amongst the fields), assigns a
coefficient to each one, and writes down the Lagrangian as the most
general linear combination of these terms.  The values of these
coefficients are then to be established by experimental measurements.
The SME ``extends'' this paradigm by relaxing the requirement that the
field combinations in the Lagrangian be Lorentz-invariant.  Since the
Lagrangian itself should still be a Lorentz scalar, the coefficients
of these new terms must have non-trivial tensor structure (rather than
being Lorentz scalars as in the original Standard Model.)  These new
coefficients (or, more accurately, their components in some reference
frame) can then in principle be measured via experiment.

While this method works well for the purposes of particle physics, it
becomes somewhat problematic when we attempt to extend it to gravity.
With a flat metric, it is legitimate to view the new Lorentz-tensor
coefficients in the SME as constants throughout spacetime.  The notion
of a constant tensor field on a flat background spacetime is
well-defined; we simply require that (for example) $\partial_a v^b =
0$.  However, once we allow for a curved background, it is no longer
so simple to find a covariantly constant non-zero vector field (i.e.,
$\nabla_a v^b = 0$); indeed, such a vector field may not even exist on
an arbitrary curved background.  Moreover, general arguments involving
the Bianchi identities \cite{Kosgrav} imply that any ``background
tensor field'' that couples directly to the curvature in the
Lagrangian must satisfy certain differential conditions; we cannot
simply write down some fixed tensor fields on our manifold and proceed
from there.

The standard way to solve the problems arising in making the metric
dynamical is to also promote the SME coefficients to dynamical fields.
As the Lorentz-violating fields are now dynamical, there is no reason
to expect them to be covariantly constant, and the geometric
consistency of the field configuration with the Bianchi identities is
automatic.  One then constructs the theory such that these fields take
on some non-zero value in the limit of no conventional matter and flat
spacetime (hence ``violating'' Lorentz symmetry by taking on a
non-invariant background value.)  The flat-spacetime SME is recovered
by constructing an effective field theory about this background, where
the metric and the Lorentz-violating fields are held fixed but the
other fields in the theory are allowed to vary.

While this promotion of Lorentz-violating coefficients to
Lorentz-violating fields solves the above problems, it does require
some care.  In particular, the requirement that the dynamics of a
Lorentz-violating vector field not fundamentally change the dynamics
of the metric restricts us to a small subclass of all conceivable
vector models \cite{VMLB}.  The resulting models have the property
that the dynamics of the field are decoupled from the dynamics of the
metric (at least at the linearized level.)  

This decoupling between the Lorentz-violating field and the metric had
been previously seen in simpler vector models known as ``bumblebee
models'' \cite{BKNG, BFKmass}.  It was further noted that the
linearized equations of such a system were (under minor auxiliary
conditions) precisely those of linearized Einstein-Maxwell theory.
The perturbations of the Lorentz-violating vector field could then be
interpreted as the photon field in such a theory.  Under such an
interpretation, the dynamics of the known long-range forces would be
the same as in conventional Einstein-Maxwell theory, up to
small Planck-suppressed deviations.

The class of models found in \cite{VMLB} included the previously-known
bumblebee models, and so were dubbed ``generalized bumblebee models''.
However, it was noted in that paper that attempting to extend the
``bumblebee photon'' interpretation to these generalized models could
lead to readily observable effects.  Specifically, in a generalized
bumblebee model the metric perturbations and conventional matter will
``see'' a different metric than the photon field will; in other words,
the ``speed of light'' would differ from the ``speed of gravity'' and
the limiting speed of conventional matter.

The present work elaborates on the above speculation.  Specifically,
we will derive the observable consequences of the ``generalized
bumblebee photon'' theory, and place bounds on the parameters of the
underlying Lorentz-violating vector field.  Section \ref{GBMrevsec}
reviews the derivation of generalized bumblebee models and describes
the ``photon interpretation'' mentioned above.  Section
\ref{bumbphotsec} derives the possible experimental signatures for
such theories, both in the context of the SME and in terms of particle
kinematics.  Finally, Section \ref{expsec} examines the current
experimental and observational bounds on Lorentz-symmetry violation in
the photon sector, and derives bounds on the parameters of the
underlying Lorentz-violating field.

We will use units in which $\hbar =8 \pi G = 1$ throughout.  We will also
set $c = 1$ in Section \ref{GBMrevsec};  however, to avoid confusion
between the various limiting speeds in subsequent sections, we will
explicitly include all such speeds in our equations.  Sign conventions
concerning the metric and the curvature tensors are those of Wald
\cite{Wald}, except in the Appendix where we use the signature
$(+,-,-,-)$.  

\section{Generalized bumblebee models}
\label{GBMrevsec}

If we limit ourselves to theories of second differential order
containing a single vector field $B_a$ along with the metric $g_{ab}$,
the most general model of dynamical Lorentz symmetry breaking has the
action
\begin{multline}
  \label{toogen}
  S = \int \dif^4 x \sqrt{-g} (R + \mathcal{J}^{abcd} R_{abcd} \\ +
  \mathcal{K}^{abcd} \nabla_a B_b \nabla_c B_d - V(B^2) )
\end{multline}
where $R$ is the Ricci scalar derived from $g_{ab}$, $R_{abcd}$ the
Riemann tensor, and $\mathcal{J}^{abcd}$ and $\mathcal{K}^{abcd}$ are
arbitrary tensors constructed locally out of $B_a$ and $g_{ab}$.  The
potential $V(B^2)$ is taken to vanish and to be minimized at some
non-zero value of its argument.  Under these assumptions, any field
configuration with
\begin{equation}
  \label{flatapprox}
  g_{ab} = \eta_{ab} \text{ and } B_a = \bb_a,
\end{equation}
where $\bb_a$ is a constant non-zero vector field with $V(\bb^2) =
V'(\bb^2) = 0$, is the ``natural'' solution of the equations of
motion.  This non-zero vector field then provides a ``preferred
direction'' in spacetime.

While a Lagrangian of the form \eqref{toogen} is indeed the most
general form for the Lagrangian, it was shown \cite{VMLB} the
equations of motion derived from a completely arbitrary Lagrangian
have certain less-than-desirable properties.  In particular, when
varying the kinetic term for the vector field $\mathcal{K}^{abcd}
\nabla_a B_b \nabla_c B_d$, we find that it gives rise to
second derivatives of $B_a$ in the Einstein equation, and that these
terms cannot in general be eliminated via the vector equation of
motion.  We are thus left with a situation in which the dynamics of
the vector field are inherently coupled to those of the metric.

However, a certain class of vector models do not exhibit this
coupling.  In particular, if the kinetic term for the vector field is
of the form
\begin{equation}
  \label{psMaxdef}
  \mathcal{K}^{abcd} \nabla_a B_b \nabla_c B_d = -
  \zeta \tg^{ab} \tg^{cd} F_{ac} F_{bd},
\end{equation}
where
\begin{equation}
  \label{tgdef}
  \tg^{ab} = g^{ab} + \beta B^a B^b
\end{equation}
and $F_{ab} = 2 \nabla_{[a} B_{b]}$, then the equations of motion for
the vector field and the metric decouple.\footnote{Although we will
  take $\zeta$ and $\beta$ to be constants for most of the paper, it
  is possible that they might themselves be functions of $B^2$.  If
  this is the case, the decoupling still holds at the level of the
  linearized equations, with $\zeta$ and $\beta$ being replaced by
  $\zeta(\bb^2)$ and $\beta(\bb^2)$.}  (The negative sign is chosen
for agreement with convention; $\zeta$ may be positive or negative.)
This decoupling is not terribly surprising when one remembers that the
field strength $F_{ab}$ is proportional to the exterior derivative of
a one-form, and thus is independent of the derivative operator;
varying the metric (and its associated covariant derivative operator)
therefore does not give rise to any terms containing the second
derivatives of the vector field as it does in the general case.  (See
\S4 of \cite{IsenNest} for further discussion.)

For the remainder of the paper, we will restrict our attention to
theories with ``pseudo-Maxwell'' kinetic terms of this type.  We will
also take $\mathcal{J}^{abcd}$ to vanish;  the primary effect of such
terms in pseudo-Maxwell vector theories is to modify the ``effective
Einstein equation'' \cite{BK, VMLB}, but gravitational effects are not
the primary focus of this paper.  Our models will thus be those
derived from an action of the form
\begin{equation}
  \label{ourmodels}
  S = \int \dif^4 x \sqrt{-g} \left( R - \zeta \tg^{ab} \tg^{cd}
    F_{ac} F_{bd} - V(B^2) \right).
\end{equation}

\subsection{Linearized equations}

Varying $g^{ab}$ and $B_a$ in the action \eqref{ourmodels}, we find
that the full equations of motion are of the form
\begin{multline}
  \label{fullEin}
  G_{ab} = 2 \zeta \Big( F_{ac} F_{bd} \tg^{cd} - \frac{1}{4} g_{ab}
    F_{cd} F_{ef} \tg^{ce} \tg^{df}  \\ + 2 \beta B_{(a} F_{b)c} F_{de}
    \tg^{ce} B^d \Big) \\ + V'(B^2) B_a B_b - \frac{1}{2} g_{ab} V(B^2)
\end{multline}
and
\begin{equation}
  \label{fullvec}
  \nabla_c \left( \tg^{ad} \tg^{cb} F_{bd} \right) + \beta F^{a} {}_b
  F_{cd} \tg^{bc} B^d = \frac{1}{2 \zeta} V'(B^2) B^a.
\end{equation}

We now linearize these equations about the background described above,
i.e., we let
\begin{equation}
  \label{hdef}
  g_{ab} = \eta_{ab} + h_{ab}
\end{equation}
and
\begin{equation}
  \label{Adef}
  B_a = \bb_a + A_a,
\end{equation}
where $\bb_a$ is a constant vector field on Minkowski spacetime and
$h_{ab}$ and $A_a$ are considered to be ``small''.  Requiring that
this field configuration be a solution when $h_{ab}$ and $A_a$ vanish
implies that $V(\bb^2) = V'(\bb^2) = 0$, as noted above.  If we then
linearize the equation \eqref{fullEin} about this background, we
obtain the linearized Einstein equation
\begin{equation}
  \label{linEin}
  \delta G_{ab} = V''(\bb^2) \bb_a \bb_b \delta(B^2),
\end{equation}
where $\delta G_{ab}$ is the linearized Einstein tensor (in terms of
derivatives of $h_{ab}$) and $\delta(B^2)$ is the linearized variation
in the norm of $B_a$,
\begin{equation}
  \label{deltB}
  \delta(B^2) = 2 \bb^a A_a - \bb^a \bb^b h_{ab}.
\end{equation}
Linearizing the vector equation of motion \eqref{fullvec}, meanwhile,
yields 
\begin{equation}
  \label{linvec}
  \te^{ad} \te^{bc} \partial_b ( \partial_c A_d - \partial_d A_c ) =
  \frac{1}{2 \zeta} V''(\bb^2) \bb^a \delta(B^2),
\end{equation}
where $\te^{ab}$ is the background value of $\tilde{g}^{ab}$ \eqref{tgdef}, the ``effective metric'' for $B^a$:
\begin{equation}
  \label{tedef}
  \te^{ab} = \eta^{ab} + \beta \bb^a \bb^b.
\end{equation}

\subsection{Charged-dust equivalence}

The linearized equations of motion \eqref{linEin} and \eqref{linvec},
though simpler than the full equations \eqref{fullEin} and
\eqref{fullvec}, are still somewhat complex.  To gain some intuition
about their solutions, let us first consider the case of the theory in
which $\beta = 0$; in this case, the linearized vector equation
becomes
\begin{equation}
  \label{b0vec}
  \partial^b f_{ba} = \frac{1}{2 \zeta}
  V''(\bb^2) \bb_a \delta(B^2)
\end{equation}
where $f_{ab} = 2 \partial_{[a} A_{b]}$.
It was noted by Bluhm, Fung, and Kosteleck\'{y} \cite{BKNG, BFKmass} that
there is a one-to-one correspondence between solutions of these
equations with $\delta(B^2) = 0$ (in a particular gauge) and solutions
of conventional linearized Einstein-Maxwell theory (in a particular
gauge.)  Specifically, if we apply an infinitesimal diffeomorphism
(parametrized by a vector field $\xi^a$) to our background field
configuration $g_{ab} = \eta_{ab}$ and $B_a = \bb_a$, these fields
transform as
\begin{subequations}
\begin{equation}
\eta_{ab} \to \eta_{ab} + 2 \partial_{(a} \xi_{b)}
\end{equation}
and
\begin{equation}
\bb_a \to \bb_a + \bb^b \partial_a \xi_b,
\end{equation}
\end{subequations}
since $\partial_a \bb_b = 0$.  The transformations
\begin{subequations}
\begin{equation}
h_{ab} \to h'_{ab} = h_{ab} + 2 \partial_{(a} \xi_{b)}  
\end{equation}
\begin{equation}
A_a \to A'_a = A_a + \bb^b \partial_a \xi_b
\end{equation}
\end{subequations}
are therefore gauge transformations and do not affect any physical
quantities.  In particular, for a solution of \eqref{linEin} and
\eqref{b0vec} with $\delta(B^2) = 0$, we can apply a gauge
transformation with $\xi^a$ chosen such that
\begin{equation}
\bb^a \partial_{(a} \xi_{b)} = - \frac{1}{2} \bb^a h_{ab},
\end{equation}
thereby putting $h_{ab}$ in the axial gauge (i.e., $\bb^a h'_{ab} =
0$.)  Importantly, under such a gauge transformation, we will also
have
\begin{equation}
\bb^a A'_a = \bb^a A_a + \bb^a \bb^b \partial_a \xi_b = \bb^a A_a -
\frac{1}{2} \bb^a \bb^b h_{ab} = 0,
\end{equation}
since we are assuming that $\delta(B^2)$, as given in \eqref{deltB},
vanishes.  Thus, putting $h_{ab}$ in axial gauge automatically also
puts $A_a$ in axial gauge if $\delta(B^2) = 0$.  Moreover, in this
case the equations of motion \eqref{linEin} and \eqref{b0vec} are
simply the source-free Einstein and Maxwell equations.  Thus, every
solution of \eqref{linEin} and \eqref{b0vec} for which $\delta(B^2) =
0$ can be mapped to a solution of conventional Einstein-Maxwell theory
for which both the metric perturbation and the vector field are in
axial gauge.  This mapping can also be seen to go the other way (again
up to gauge transformations):  given a solution of conventional
source-free Einstein-Maxwell theory, apply gauge transformations to
both $h_{ab}$ and $A_a$ such that they are in axial gauge with respect
to $\bb_a$.  This gauge transformation guarantees that $\delta(B^2)$
vanishes, and thus this field configuration is also a solution of
\eqref{linEin} and \eqref{b0vec} with $\delta(B^2) = 0$.

As it turns out, this mapping can be extended to the case where
$\delta(B^2) \neq 0$, at least in the case where $\bb_a$ is timelike.
Let us suggestively define\footnote{Note that the quantities $\rho_m$
  and $\rho_e$ are proportional to the quantity $\beta$ defined in
  Eqn.~(68) of \cite{BFKmass}.}
\begin{equation}
  \label{rhomdef}
  \rho_m = - V''(\bb^2) \bb^2 \delta(B^2),
\end{equation}
\begin{equation}
  \label{rhoedef}
  \rho_e = \pm \frac{V''(\bb^2) \sqrt{-\bb^2}}{2 \zeta} \delta (B^2),
\end{equation}
 and 
\begin{equation}
  \label{udef}
  u_a = \pm \frac{\bb_a}{\sqrt{-\bb^2}}.
\end{equation}
The signs of $\rho_e$ and $u_a$ are chosen to be positive if $\bb^a$
is future-directed and negative if it is past-directed; in other
words, $u^a$ is defined to be future-directed.  Rewriting
\eqref{linEin} and \eqref{b0vec} in terms of these quantities, the
equations become
\begin{subequations}
\begin{equation}
  \delta G_{ab} = \rho_m u_a u_b
\end{equation}
\begin{equation}
  \partial^b f_{ba} = \rho_e u_a
\end{equation}
\end{subequations}
which are easily recognizable as the equations of motion for the
perturbed metric and the vector field $A_a$ in the presence of charged
dust.  By construction, $u_a$ is a unit, future-directed timelike
vector.  The charge-to-mass ratio of the dust is constant, and is
given by
\begin{equation} 
  \label{cmratio}
  \rho_e / \rho_m = \mp \frac{1}{2 \zeta \sqrt{-\bb^2}}.
\end{equation}
By applying the Bianchi identity to \eqref{b0vec}, we obtain
\begin{equation}
\bb^a \partial_a (\delta(B^2) ) = 0,
\end{equation}
which guarantees that $\rho_m$ and $\rho_e$ are constants along the
worldlines parametrized by $u^a$.  

We therefore conclude that any fields $h_{ab}$ and $A_a$ satisfying
\eqref{linEin} and \eqref{b0vec} can be mapped to a solution of
conventional Einstein-Maxwell theory with a charged dust source, where
the dust moves along the worldlines parametrized by $\bb^a$, and its
mass density and charge density are given by \eqref{rhomdef} and
\eqref{rhoedef} respectively. As in the case of vanishing
$\delta(B^2)$, this correspondence goes the other way as well.
Suppose we have a solution $\{h_{ab}, A_a\}$ of the linearized
Einstein-Maxwell equations with a charged-dust source with mass
density $\rho_m$ and charge density $\rho_e$, with $\rho_m$ and
$\rho_e$ satisfying \eqref{cmratio}.  We can perform a gauge
transformation on the Maxwell field, $A_a \to A_a + \partial_a
\lambda$, with $\lambda$ satisfying
\begin{equation}
  \bb^a \partial_a \lambda = - \frac{\rho_m}{2 V''(\bb^2) \bb^2} -
  \bb^a A_a + \frac{1}{2} \bb^a \bb^b h_{ab}
\end{equation}
(This does not uniquely determine $\lambda$, of course, but we only
require $\lambda$ to exist.)  Under this gauge transformation, the
fields $h_{ab}$ and $A_a$ will satisfy
\begin{equation}
  \delta(B^2) = 2 \bb^a A_a - \bb^a \bb^b h_{ab} = - \frac{\rho_m}{
    V''(\bb^2) \bb^2}.
\end{equation}
We can then see that in this gauge the fields $h_{ab}$ and $A_a$
satisfy our original equations \eqref{linEin} and \eqref{b0vec}.  This
correspondence is easily seen to agree with the original
correspondence \cite{BKNG, BFKmass} in the case where $\delta(B^2) =
0$.

This correspondence, between solutions of our linearized equations
\eqref{linEin} and \eqref{b0vec} and those of
Einstein-Maxwell-charged-dust systems, can be then used to gain some
intuition about the behaviour of our system.\footnote{Our
  correspondence also seems to work in the case of spacelike $\bb_a$.
  However, in this case the ``dust'' sources will be moving along
  spacelike worldlines, a situation of which it is less common to have
  an intuitional understanding.}  In particular, this correspondence
justifies the tactic (used in \cite{BKNG, BFKmass}) of simply setting
the $\delta(B^2)$ term to zero.  One might have been concerned that
this set of solutions was unstable, in the sense that a solution with
$\delta(B^2)$ initially small but non-zero might evolve to a solution
with large $\delta(B^2)$.  This new correspondence shows that this is
not the case, since sufficiently small $\delta(B^2)$ on the
``bumblebee'' side corresponds to small sources on the
Einstein-Maxwell side, and a solution of the conventional
Einstein-Maxwell equations with a small source will be ``close'' (in
an appropriate sense) to a solution of the Einstein-Maxwell equations
with no sources.  From here on, we will assume that $\delta(B^2)$ is
negligible unless otherwise stated.  

A similar correspondence was noted by Jacobson and Mattingly
\cite{einaeth} in their studies of ``Einstein-{\ae}ther theory.''  In
this case, however, the vector field they were examining served a dual
purpose as both the vector potential and the dust worldlines; this
implied, in particular, that their ``dust'' was dynamical rather than
a fixed background source.  Since these two vectors are not in general
aligned in an arbitrary Einstein-Maxwell-charged dust system, the
correspondence found in \cite{einaeth} was therefore not one-to-one
(even after taking gauge transformations into account.)  Since our
correspondence uses the fixed background as a ``source'' for the
linearized perturbations, it does not run into this difficulty;  any
solution of the linearized bumblebee equations can be
gauge-transformed into a solution of the Einstein-Maxwell equations
with a charged-dust source, and vice versa.  

Finally, recall that all of the analysis in this subsection has been
done assuming that the constant $\beta$ vanishes.  The above analysis
changes in two main ways if $\beta \neq 0$, one less important and one
more important.  The first is that the charge-to-mass ratio of the
dust in the above correspondence changes.  The linearized Maxwell
equation \eqref{linvec} in this case becomes
\begin{equation}
  \te^{bc} \partial_b f_{ca} = \frac{V''(\bb^2) \sqrt{-\bb^2}}{2
    \zeta (1 + \beta \bb^2)} \delta(B^2) u_a
\end{equation}
with $u_a$ defined as in \eqref{udef}.  (To see this, multiply by the
tensor $\te_{ab}$ defined such that $\te_{ab} \te^{bc} = \delta_a
{}^c$.)  Thus, the ``charge density'' defined in \eqref{rhoedef} is
multiplied by a factor of $(1 + \beta \bb^2)^{-1}$ when we pass to the
general case of non-vanishing $\beta$.\footnote{We are of course
  assuming here that $\beta \bb^2 \neq -1$.  In the case where $\beta
  \bb^2 = -1$, the inverse metric defined in \eqref{tedef} becomes
  degenerate, and \eqref{linvec} cannot be viewed as an evolution
  equation.  We will assume hereafter that $\beta$ and
  $\bb^2$ are chosen such that $\eta^{ab}$ and $\te^{ab}$ have the
  same signature.} The rest of the above argument holds, however; in
particular, we are still justified in assuming $\delta(B^2)$ to be
negligible.

More importantly, however, when $\beta \neq 0$ the vector
perturbations $A_a$ will not propagate with the same velocity as those
of the metric.  Instead, the metric perturbations will propagate along
the light-cones of the usual flat metric $\eta^{ab}$, while the vector
perturbations will propagate along the light-cones of the ``bumblebee
metric'' $\te^{ab}$.  Assuming that any matter sources are minimally
coupled to the ``Einstein metric'' $g_{ab}$ used in the action
\eqref{ourmodels}, and that their kinetic terms are not directly
coupled to $B^a$, this also implies that the limiting speed of
conventional matter will be different from the limiting speed of the
vector perturbations.  The observational consequences of this fact
will be explored in the next section.

\section{Lorentz-violating photons}
\label{bumbphotsec}

\subsection{Bumblebee photon theories}
\label{nowitsaphoton}

We found in the last section that linearized solutions of the
equations \eqref{fullEin} and \eqref{fullvec} (about a background
where the metric is flat and the vector is non-zero) can be taken to
satisfy the equations
\begin{subequations}
  \begin{equation}
    \delta G_{ab} = 0
  \end{equation}
and
  \begin{equation}
    \te^{bc} \partial_b f_{ca} = 0.
  \end{equation}
\end{subequations}
From the perspective of particle physics, these are massless fields
(or more precisely, Nambu-Goldstone modes arising from a spontaneously
broken symmetry.)  One can envision a number of distinct possibilities
concerning the effects of such fields on the theory:
\begin{itemize}
  \item \emph{The field $B_a$ does not directly couple to conventional
      matter.}  In this case, we would not have detected its effects
    in particle experiments.  The effects of the Lorentz-violating
    field might still be observable via gravitational effects
    \cite{BK}, but would not give rise to forces between particles of
    ``conventional'' matter.

  \item \emph{The Nambu-Goldstone modes are ``eaten'' by another field
      via a Higgs mechanism.}  This turns out to be impossible
    \cite{KosSam} in the context of spacetime with a Riemann metric,
    though it is possible in Riemann-Cartan spacetimes with a
    dynamical torsion field \cite{BKNG}.  We will not consider this
    possibility further here.

  \item \emph{The massless field $A_a$ couples directly to
      conventional matter, giving rise to a long-range ``fifth
      force''.}  For example, the field $A^a$ could conceivably couple
    to leptons but not quarks (or vice versa), it could couple
    differently to first-generation particles than to
    second-generation particles; it could couple only to strange
    quarks; and so on.  The large number of experimental signatures
    that conceivably could arise in such scenarios are, unfortunately,
    outside the scope of this paper;  models along these lines have
    been explored in \cite{AHspinforce, KTgrava}.

  \item \emph{The massless field $A_a$ couples directly to
      conventional matter, and can be identified with a known force.}
    The obvious candidate here (as may have been telegraphed by the
    choice of notation) would be the photon field \cite{BKNG,
      BFKmass}.  This interpretation will be the focus of the rest of
    this work.
\end{itemize}

One might ask whether it is self-consistent to demand that $\eta^{ab}$
serve as the ``conventional matter metric'' while simultaneously
requiring that the bumblebee perturbations $A_a$ serve as the photon.
This self-consistency can be shown by examining the possible couplings
between $B^a$ and the fermion fields in the theory.  Suppose we have a
fermion field $\psi$ appearing in the Lagrangian with its standard
kinetic term and with $B_a$ coupling to its current:
\begin{align}
  \label{ferminta}
  \mathcal{L}_\psi &= \frac{i}{2} \bar{\psi} \gamma^a \doubarrow{\partial}_a \psi
  + q B_a \bar{\psi} \gamma^a \psi \nonumber \\
  &=  \frac{i}{2} \bar{\psi} \gamma^a \doubarrow{\partial}_a \psi + q \bb_a
  \bar{\psi} \gamma^a \psi + q A_a \bar{\psi} \gamma^a 
  \psi.
\end{align} 
We can then see that the decomposition of $B_a$ into a background
field plus a perturbation (identified as the photon) leads to the
usual interaction term $q A_a \bar{\psi} \gamma^a \psi$ between the
bumblebee photon and the fermion.  The second term on the right-hand
side of \eqref{ferminta}, meanwhile, can be interpreted in the
language of the Standard Model Extension (SME) \cite{CKSME} as a
Lorentz-violating coefficient $a_a = q \bb_a$.  Through a redefinition
of the spinor phases, the coefficients $a_a$ in the SME can be made to
vanish in flat spacetime \cite{CPTviol}.  Thus, a term of the form
\eqref{ferminta} would give rise to a conventional photon-fermion
interaction, without other observable effects in the fermion sector.

We could also envision having the fermion interact with the bumblebee
field via a derivative interaction:
\begin{equation}
  \label{fermintc}
  \mathcal{L}_\psi = \frac{i}{2} \bar{\psi} \gamma^a
  \doubarrow{\partial}_a 
  \psi + \frac{i}{2} q_c B^a B^b \bar{\psi} \gamma_a
  \doubarrow{\partial}_b \psi
\end{equation}
where $q_c$ is a coupling coefficient.  In the language of the SME,
such a term would give rise to a $c_{ab}$ coefficient for the fermion
field $\psi$.\footnote{In principle, we could also couple $B_a$ to the
  axial fermion current $\bar{\psi} \gamma_5 \gamma^a \psi$ or to a
  term of the form $i \bar{\psi} \gamma_5 \gamma_a
  \doubarrow{\partial}_b \psi$; such terms would give rise to $b_a$
  and $d_{ab}$ coefficients in the SME, respectively.  In this work,
  we will assume these vanish.}  It is precisely such a $c_{ab}$ term
that would cause the ``effective fermion metric'' to differ from
$\eta^{ab}$.\footnote{Such a term would also give rise to
  momentum-dependent fermion-fermion-photon vertices, as well as
  two-photon-two-fermion vertices; however, such terms would be
  nonrenormalizable, and therefore would be highly suppressed at low
  energies.}  If we consider the bumblebee field as taking on its
fixed background value, the above Lagrangian \eqref{fermintc} can be
rewritten as
\begin{equation}
  \mathcal{L}_\psi = \frac{i}{2} \breve{\eta}^{ab} \bar{\psi} \gamma_a
  \doubarrow{\partial}_b \psi
\end{equation}
where $\breve{\eta}^{ab} = \eta^{ab} + q_c \bb^a \bb^b$.  We could
equally well define $\breve{\eta}^{ab}$ to be our ``fundamental
metric'' instead of $\eta^{ab}$; this essentially amounts to a
rescaling of the coordinates \cite{LehKos}.  We would thus have a
theory in which the electrons propagate with respect to the
``fundamental metric'' $\breve{\eta}^{ab}$, the photons propagate with
respect to
\begin{equation}
  \te^{ab} = \breve{\eta}^{ab} + (\beta - q_c) \bb^a \bb^b
\end{equation}
and the metric perturbations propagate with respect to
\begin{equation}
  \eta^{ab} = \breve{\eta}^{ab} - q_c \bb^a \bb^c.
\end{equation}
We can then see that a theory with a non-vanishing $q_c$ is physically
equivalent to a theory with $q_c \to 0$, $\beta \to \beta -
q_c$, and a ``distorted metric'' for the metric perturbations.  As the
remainder of the paper will not be concerned with the metric
perturbations, we will therefore assume that $q_c$ has been set to
zero in this way, and we will use $\eta^{ab}$ to denote the ``matter
metric''.

\subsection{SME coefficients}

If $A_a$ is to be interpreted as the photon field in our theory, we
immediately note an important experimental consequence of this fact:
the photon does not propagate along the null cones of the conventional
matter metric $\eta^{ab}$, but rather along those of the distorted
metric $\te^{ab}$ defined in \eqref{tedef}.  This distortion will, in
principle, be experimentally detectable.  The potential effects of a
background geometric structure on the propagation of photons were
explored in detail by Kosteleck\'y and Mewes \cite{KMsig}.  One starts
with a photon Lagrangian of the form
\begin{equation}
  \label{LVphot}
  \mathcal{L} = F_{ab} F^{ab} + (k_F)^{abcd} F_{ab} F_{cd}
\end{equation} 
(up to an overall normalization), where $(k_F)^{abcd} =
(k_F)^{[ab][cd]}$ is symmetric under the exchange $\{ab\}
\leftrightarrow \{cd\}$ and has vanishing double trace (i.e.,
$(k_F)^{ab} {}_{ab} = 0$.)  The tensor $(k_F)^{abcd}$ can then be
decomposed into various ``electric'' and ``magnetic'' parts that
determine the electric and magnetic susceptibility of free space, as
well as vacuum birefringence effects.  In our case, the effective
flat-space Lagrangian for $A_a$ is given by
\begin{equation}
  \mathcal{L} = \te^{ac} \te^{bd} F_{ab} F_{cd}
\end{equation}
which corresponds to a $(k_F)^{abcd}$ tensor of
\begin{equation}
  \label{kFdef}
  (k_F)^{abcd} = - \frac{\beta}{1 + \frac{\beta}{2} \bb^2} \left(
    \frac{\bb^2}{2} \eta^{c[a} \eta^{b]d} + 2 \bb^{[a} \eta^{b][c}
    \bb^{d]} \right).
\end{equation}
(The factor in the denominator arises from factoring out the overall
normalization mentioned above.)  For the purposes of comparison with
experiment, the components of $(k_F)^{abcd}$ can be decomposed into
four spatial matrices $\tk_{e \pm}$ and $\tk_{o \pm}$ and a trace
component $\tk_\text{tr}$, as defined in Section II B of
\cite{KMsig}.\footnote{Note that we have also implicitly defined a
  reference frame in defining these as ``spatial'' matrices. In what
  follows, we take this frame to be the standard Sun-centred frame,
  where the Sun is at rest, the $Z$-axis points towards the North
  Celestial Pole, and the $X$-axis points towards the Vernal Equinox.}
In the current case, these work out to be
\begin{subequations}
\begin{equation}
  \label{kape+o-}
  (\tk_{e+})^{ij} = (\tk_{o-})^{ij} = 0,
\end{equation}
\begin{equation}
  \label{kape-def}
  (\tk_{e-})^{ij} = \tb \left( \bb^i \bb^j - \frac{1}{3} \delta^{ij}
    \vec{B}^2 \right),
\end{equation}
\begin{equation}
  \label{kapo+def}
  (\tk_{o+})^{ij} = \tb \bb^0 \epsilon^{ij} {}_k \bb^k,
\end{equation}
and
\begin{equation}
  \label{kaptrdef}
  \tk_\text{tr} = - \frac{\tb}{2} \left( (\bb^0)^2 + \frac{1}{3}
    \vec{B}^2 \right)
\end{equation}
\end{subequations}
where $\vec{B}$ denotes the spatial components of $\bb^a$, and we have
defined
\begin{equation}
  \tb \equiv \frac{ \beta}{1 + \frac{\beta}{2} \bb^2}.
\end{equation}
We can then use experimental measurements (see \cite{LVdata} and
references therein) of the components of $\tk_{e \pm}$, $\tk_{o \pm}$,
and $\tk_\text{tr}$ to place bounds on the parameters $\beta$ and
$\bb^\mu$ of our theory.

\subsection{Particle kinematics}
\label{partkin}

\begin{figure}
  \begin{center}
    \includegraphics[scale=0.9]{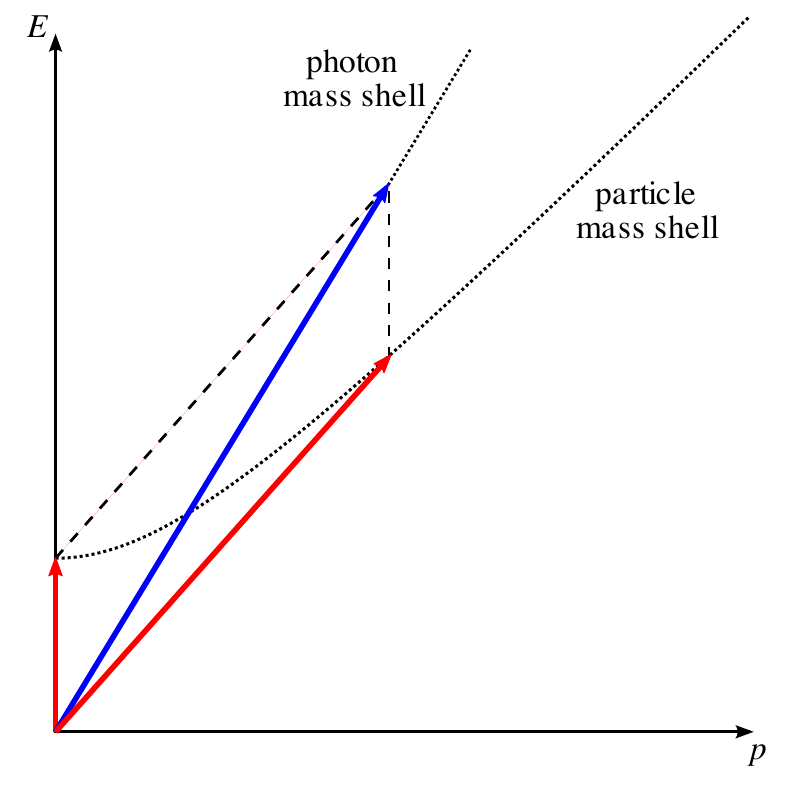} \\ (a) \\[1ex]
    \includegraphics[scale=0.9]{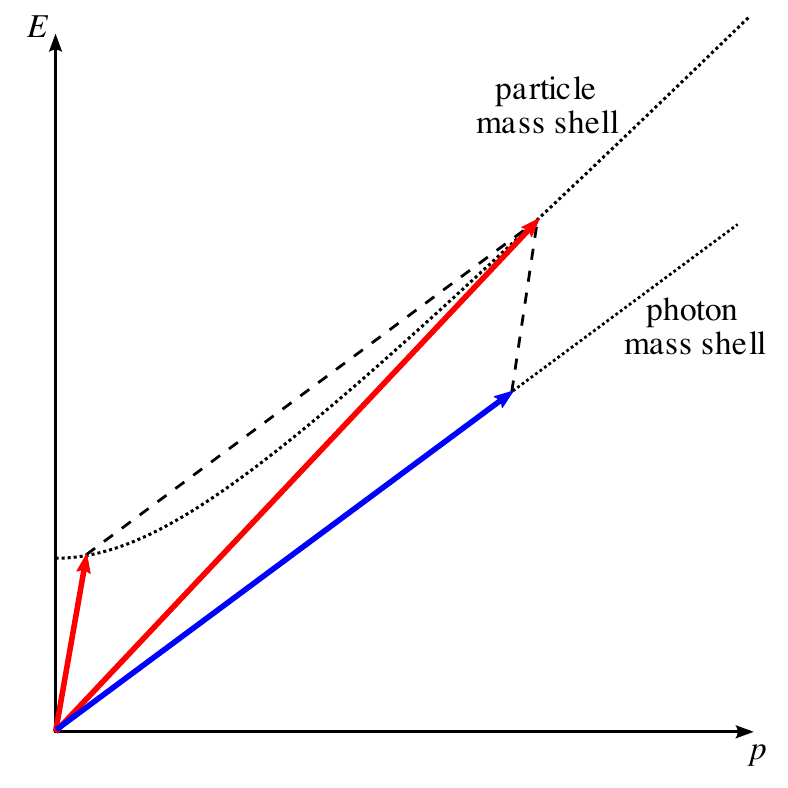} \\ (b)
  \end{center}
  \caption{Mass shells and momentum vectors for one-to-two processes
    when the limiting particle speed $c_p$ differs from the speed of
    light $c_\gamma = E_\gamma/p_\gamma$. (a) When $c_\gamma > c_p$
    (or, equivalently, $\beta >0$), a photon (blue) can decay into two
    massive charged particles 
    (red). (b) When $c_\gamma < c_p$ ($\beta < 0$), a charged 
    particle with sufficiently high energy (red) can decay to a
    charged particle of lower energy (red) and a photon (blue).  
    \label{kinfig} }
\end{figure}

In a theory in which the limiting speed of a charged particle species
is identical to the speed of the photon, it is kinematically forbidden for
a photon to decay to that particle and its antiparticle, or for the
charged particle to radiate a photon.  When the limiting speed of a
charged particle differs from the speed of light propagation, however,
such processes are kinematically allowed (see Figure \ref{kinfig}.)
More precisely, if the speed of light $c_\gamma$ in a given direction
is \emph{greater} than the limiting particle speed $c_p$ in that
direction, then photons above a certain energy can decay; if
$c_\gamma$ in a given direction is \emph{lesser} than $c_p$ in that
direction, then particles above a certain energy will undergo vacuum
\v{C}erenkov radiation.

It is important to note that each of these processes is sensitive to
$\beta$ values of only one sign.  The directions $k_a$ of bumblebee
photon propagation are those for which $\te^{ab} k_a k_b = 0$.
Denoting $k_\mu = (\omega/c_p, \vec{k})$, where $c_p$ is the limiting
speed of the particle species, we see that this ``null condition'' is
equivalent to
\begin{equation}
  \label{disp1}
  \omega^2 = c_p^2 (\vec{k}^2 + \beta (\bb^a k_a)^2)
\end{equation}
This implies that if $\beta > 0$, we will have $c_\gamma =
\omega/|\vec{k}| > c_p$, and photons of sufficiently high energy can
decay to charged particles and anti-particles.  Similarly, if $\beta <
0$, we will have $c_\gamma < c_p$, and charged particles of
sufficiently high energy will lose energy to vacuum \v{C}erenkov
radiation.\footnote{The generalized bumblebee photon model has what
  might be called a ``homogeneous'' photon dispersion relation: for
  all $\lambda >0$, if $k^a$ is a valid four-momentum for a
  propagating photon, then so is $\lambda k^a$.  Our discussion below
  can easily be extended to any model in which this is the case.  For
  models in which this does not hold (see, for example,
  \cite{ChernSimonsphoton}), the ``geometric'' arguments used below to
  find the vacuum \v{C}erenkov threshold can be adapted to analyze
  both photon decay and vacuum \v{C}erenkov processes.}

What are the threshold energies for these processes?  Denoting
$\vec{k} = k \hat{k}$, where $\hat{k}$ is a unit vector, we find that
the photon dispersion relation \eqref{disp1} above can written as
\begin{equation}
  \label{disp2}
  \omega = c_p k \left[ \frac{ \beta \bb^0 (\vec{B} \cdot \hk) +
      \sqrt{1 + 
      \beta( -(\bb^0)^2 + (\vec{B} \cdot \hk)^2 )}}{1 - \beta
    (\bb^0)^2} \right]
\end{equation} 
In the case of photon decay, the threshold photon energy $E_{d0}$ will be
that for which a photon with four-momentum $\hbar c_p k_a$ can decay
into a particle-antiparticle pair, with each particle having
four-momentum $p_a = \frac{1}{2} \hbar c_p k_a$ and rest mass $m$.
Taking the norm of $p_a$ with respect to $\te^{ab}$, we find that the
four-momentum of the each particle must satisfy
\begin{equation}
  p_a p_b \te^{ab} = p^a p_a + \beta (\bb^a p_a)^2 = -m^2 c_p^4 +
  \beta(\bb^a p_a)^2.
\end{equation}
The left-hand side of this equation vanishes (since $p_a p_b \te^{ab}
\propto k_a k_b \te^{ab}$).  Rewriting the right-hand side in terms of
$k_\mu$, and defining $n^a \equiv \sqrt{|\beta|} \bb^a$, we find that at
threshold
\begin{equation}
  ( n^0 \omega + (\vec{n}\cdot \hk) c_p k)^2 = \left( \frac{2 m
      c_p^2}{\hbar} \right)^2.
\end{equation} 
Applying the dispersion relation \eqref{disp2} then yields a threshold
energy of
\begin{equation}
  E_{d0}(\hk) = 2 m c_p^2 \left| \frac{ n^0 - (\vec{n}
      \cdot \hk) 
      \sqrt{1 - (n^0)^2 + (\vec{n} \cdot \hk)^2 }}{(n^0)^2 - (\vec{n}
      \cdot \hk)^2} \right|. 
\end{equation}
In the limit of the components of $n^\mu$ being much less than unity,
this simplifies to 
\begin{equation}
  \label{photdecthres}
  E_{d0}(\hk) \approx \frac{2 m c_p^2}{|n^0 + \vec{n} \cdot \hk |}.
\end{equation}

In the case of vacuum \v{C}erenkov radiation, the threshold energy is
best found using a geometric argument (see Figure \ref{kinfig}.)
Since momentum is conserved, the four-momentum of the photon will
``connect'' two points on the charged-particle mass shell.  The
threshold energy $E_{c0}$ for vacuum \v{C}erenkov radiation is
therefore that point on the mass shell at which the slope of the
tangent line ($\dif E_p / \dif p_p$) equals the slope of the photon
mass shell: at any energy on the charged-particle mass shell with $E_i
> E_{c0}$, we can draw a secant line with the same slope as the photon
mass shell that will intersect the charged-particle mass shell at a
lower energy $E_f < E_i$.  This will not, however, be possible for
$E_i < E_{c0}$.  Performing this calculation, we find that the
threshold energy $E_{c0}$ for a particle of mass $m$ is given by
\begin{equation}
  E_{c0}(\hk) = \frac{m c_p^2}{\sqrt{1 - c_\gamma^2(\hk)/c_p^2}}
\end{equation}
where the (direction-dependent) speed of light $c_\gamma(\hk)$ is given
dividing $\omega$ by $k$ in \eqref{disp2}.  Plugging this in, we find
that the threshold energy is given by
\begin{equation}
  E_{c0}(\hk) = m c_p^2 \left| \frac{ (\vec{n} \cdot
      \hk) - n^0
      \sqrt{1 - (n^0)^2 + (\vec{n} \cdot \hk)^2 }}{(n^0)^2 - (\vec{n}
      \cdot \hk)^2} \right|.
\end{equation}
where we have again used the rescaled vector $n^a = \sqrt{|\beta|}
\bb^a$.  If we again take the limit of small $n^\mu$, this reduces to
\begin{equation}
  E_{c0}(\hk) \approx \frac{m c_p^2}{|n^0 + \vec{n} \cdot \hk|}.
\end{equation}

In the case $\beta < 0$, the detection of a charged particle with
energy $E_p$ and mass $m$ propagating in the direction $\hk$ implies
that the \v{C}erenkov threshold energy $E_{c0}(\hk)$ for that
direction is greater than $E_p$.  Thus, we can say that such a
detection restricts the components of $n^a$ to
lie in the region
\begin{equation}
  \label{Cerineq}
  |n^0 + \vec{n} \cdot \hat{k}| < \frac{m c_p^2}{E_p}.
\end{equation}
This region is a thickened plane in $n^\mu$-space, with a total
thickness of $2m c_p^2/E_p$.  Multiple detections of charged particles
coming from different directions $\hk$ can then constrain the
parameters of our theory to a finite region of $n^\mu$-space.  Note
that the quantity appearing on the right-hand side of \eqref{Cerineq}
is simply the inverse of boost factor $\gamma$ of the charged particle
detected.

Similarly, when $\beta > 0$, the detection of a photon with energy
$E_\gamma$ propagating in the direction $\hk$ implies that the
components of $n^a = \sqrt{|\beta|} \bb^a$ satisfy
\begin{equation}
  |n^0 + \vec{n} \cdot \hat{k}| < \frac{2 m_e c_p^2}{E_\gamma}.
\end{equation}
Here, $m_e$ is the electron mass:  the $\gamma \to e^- e^+$ process
will have a lower threshold than any other photon decay channel.  Once
again, high-energy photons from various directions can then constrain
us to a finite region of $n^\mu$-space.

The rate of energy loss for vacuum \v{C}erenkov processes has been
calculated by Altschul \cite{AltschulCer}.  In particular, a charged
particle with energy just above threshold ($E_p = E_{c0} + \Delta E$,
with $\Delta E \ll E_{c0}$) will emit a photon with $E_\gamma > \Delta
E$ with a decay rate of 
\begin{equation} 
  \Gamma = \alpha Z^2 m^2 \frac{(\Delta E)^2}{2 E_p^3},
\end{equation}
where $\alpha$ is the fine structure constant, $Z$ is the particle's
charge, and $m$ is its rest mass.  The mean free path of such a
particle (assuming it to be moving with velocity $v \approx c_p$) can
then be estimated as $\ell = c_p / \Gamma$.  At higher energies, the
charged particle will mainly lose energy to larger numbers of
lower-energy photons, rather than a single photon that brings it below
threshold.  This process causes the energy to decrease even more
rapidly; we should therefore view the estimate $\ell$ above as an
upper bound on the mean free path for a particle with energy $E_p$.  

In the case of photon decay, an exact expression for the lifetime of
the photon is not yet known.  However, we can estimate (see the
Appendix) that the decay rate for photons above threshold energy
$E_{d0}$ will be on the order of magnitude of
\begin{equation}
  \Gamma \sim \alpha \beta \mathcal{B}^2 E_\gamma \sqrt{1 -
    \frac{E_{d0}^2}{E_\gamma^2}}, 
  \label{lifetimeest}
\end{equation}
where $\alpha$ is the fine structure constant and $\mathcal{B}$ is a
quantity of the same order of magnitude as the components of $\bb^a$.
Using this estimate, one can then calculate a mean free path for
photons as in the case of vacuum \v{C}erenkov radiation.  Roughly
speaking, a photon well above threshold will have a mean free path of
order $(\alpha \beta \mathcal{B}^2)^{-1}$ times its Compton
wavelength; if the photon is only barely above threshold, with energy
$E_\gamma = (1+\epsilon) E_{d0}$, its mean free path (in Compton
wavelengths) is reduced by a factor of approximately $\sqrt{\epsilon}$.

\section{Experimental constraints}
\label{expsec}

In general, the most stringent limits on the components of
$(k_F)^{abcd}$ are those that arise from searches for vacuum
birefringence \cite{LVdata}.  However, the vanishing of the matrices
$(\tk_{e+})^{ij}$ and $(\tk_{o-})^{ij}$ \eqref{kape+o-} implies that
in our case, bumblebee photons do not experience vacuum birefringence.
We must thus turn to other experimental means of searching for Lorentz
violation in the photon sector.  In the following subsections, we will
discuss limits arising from rotating electromagnetic resonator
experiments, particle accelerator experiments, and cosmic-ray
observations.

\subsection{Resonator experiments}
\label{resonsec}

If the Maxwell field is Lorentz-invariant, the frequencies of its
modes in a resonant cavity will be independent of the cavity's
orientation in space.  However, if the photon field propagates at
different speeds in different directions, it is not hard to see that
the resonant frequencies of the cavity can change if the cavity's
orientation changes: this frequency depends on the ``speed of light''
in the cavity, and this speed is direction-dependent in our model if
$\beta \neq 0$ and $\vec{B} \neq 0$.  The magnitude of this frequency
shift for a given set of matrices $\tk_{e\pm}$ and $\tk_{o\pm}$ was
calculated for a general cavity mode and geometry in \cite{KMsig}.  In
practise, this frequency shift is usually measured by setting up two
identical cavities, oriented at right angles to each other, and
rotating these two cavities together.  A difference in the speed of
light within the plane of rotation would then show up as a ``beat''
between the frequencies of the two cavities, modulating at twice the
frequency of rotation.  By looking for these ``beats'' at various
points in the Earth's rotation and revolution, all eight independent
components of the SME matrices $(\tk_{e-})^{ij}$ and $(\tk_{o+})^{ij}$
can in principle be measured.

Sensitive measurements of the components of $(\tk_{e-})^{ij}$ and
$(\tk_{o+})^{ij}$ have been performed by Herrmann \textit{et
  al.}~\cite{HerrCPT} and by Eisele, Nevsky and Schiller \cite{ENS}.
Both groups have bounded the components of $(\tk_{e-})^{ij}$ to be
$\mathcal{O}(10^{-17})$ or less, and those of $(\tk_{o+})^{ij}$ to be
$\mathcal{O}(10^{-13})$ or less.  We can translate these measured
bounds into a rough estimate of the bounds on our parameters $\beta$
and $\bb^\mu$; given the dependencies given in \eqref{kape-def} and
\eqref{kapo+def}, we would expect the magnitude of $n^0 =
\sqrt{|\beta|} \bb^0$ to be bounded below approximately $3 \times
10^{-5}$, and the components of $\vec{n}$ to be approximately $3
\times 10^{-9}$ or less.\footnote{Note that the parameters of our
  model, as defined, are degenerate; by rescaling our definition of
  $\bb^a$ as above, we can set $\beta$ to $\pm 1$.  Our physical
  parameter space is thus four-dimensional, with an additional
  discrete parameter (the sign of $\beta$.)}

It is important to note, however, that both of the above mentioned
groups \cite{HerrCPT,ENS} derived their respective bounds from their
experimental data under the assumption that all eight independent
components of $(\tk_{e-})^{ij}$ and $(\tk_{o+})^{ij}$ could be varied
independently.  In our model, this is not the case; rather, as noted
above, we have a four-dimensional parameter space (along with an
additional discrete parameter).  A more thorough analysis should thus
involve a (non-linear) regression on this parameter space, starting
from the experimental data.

\begin{table}
  \begin{tabular}{c| r@{.}l | r@{.}l | c | c }
    & \multicolumn{4}{c|}{Best Fit}  & \multicolumn{2}{c}{$1\sigma$
      Conf. Intervals} \\ 
    & \multicolumn{2}{c}{$\beta > 0$} & 
    \multicolumn{2}{c|}{$\beta < 0$} &
    \multicolumn{1}{c}{$\beta >0$} & $\beta <  0$ \\ \hline  
    $n^0$ & $0$&$69$ & $-1$&$61$ & $[-2.16,3.54]$ & $[-4.43,1.21]$ \\
    $n^X$ & $4$&$00$ & $-2$&$06$& $[0.50,6.26]$ & $[-4.21, 0.52]$ \\
    $n^Y$ & $1$&$30$ & $3$&$61$& $[-1.57, 3.83]$ & $[0.07, 5.76]$ \\
    $n^Z$ & \hspace{1ex} $-2$&$39$ \hspace{2ex} & \hspace{1ex}
    $-1$&$46$ \hspace{2ex} & $[-5.29, 0.52]$ & $[-3.88, 0.96]$ 
  \end{tabular}
  \caption{\label{resontable} Best fits and approximate $1\sigma$
    confidence intervals for $n^0$ (times $10^{-4}$) and $\vec{n}$
    (times $10^{-8}$) in the standard Sun-centred frame, derived from
    the data of Stanwix \textit{et al.}~\cite{Stan06}.}
\end{table}

To perform such an analysis, we turn to the data of Stanwix \textit{et
  al.}  In Table I of \cite{Stan06}, the time-variation of the
above-mentioned ``beat'' amplitudes are given.  Applying a non-linear
regression to these amplitudes gives the results shown in Table
\ref{resontable} for the components $n^\mu$ in the Sun-centred frame.
The best fit is found to occur for $\beta > 0$; however, a region of
parameter space with $\beta < 0$ also falls in the overall $1 \sigma$
confidence region.  The approximate dimensions of the one-sigma confidence
contour in both the $\beta >0$ and $\beta < 0$ regions of parameter
space are also given in Table \ref{resontable}.  The point $n^\mu =
(0,0,0,0)$ lies approximately on the 75\% confidence contour.  While
this might seem suggestive of a non-zero Lorentz-violating effect, such
a confidence level can hardly be thought of as conclusive, especially
considering that Stanwix \textit{et al.}~viewed as spurious the $2
\sigma$ and $3 \sigma$ signals found via their original analysis (with
SME coefficients assumed to be independent.)  We must therefore
conclude that resonator experiments have not yet observed a signal
compatible with the generalized bumblebee model.

\subsection{Accelerator physics}

As noted above, vacuum \v{C}erenkov radiation and photon decay, though
kinematically forbidden in conventional theories, are both allowed
above a certain energy threshold in the presence of Lorentz invariance
in the photon sector.  In the bumblebee photon case, this threshold
will decrease as the components of the rescaled vector $n^a =
\sqrt{|\beta|} \bb^a$ increase.  Direct observations of high-energy
particles can therefore help us constrain our theory.

An analysis along these lines has been performed by Hohensee
\textit{et al.} \cite{Hohensee}, through analysis of the operation of
the LEP experiment and the Tevatron.  In the case of vacuum
\v{C}erenkov radiation, they note that a threshold energy $E_{c0}$
more than a few MeV below the electron and positron beam energies at
LEP ($E_\text{LEP} = 104.5$ GeV) would have caused beam energy losses
significant enough to be immediately apparent.  In our case, this
implies that\footnote{The propagation direction $\hk$ is, of course,
  not a constant for the electrons in a circular accelerator.
  However, the independent bounds on $\vec{n}$ from resonator
  experiments make this consideration moot.}
\begin{equation}
  |n^0 + \vec{n} \cdot \hk| < \frac{m_e}{E_\text{LEP}} \approx 4.9
  \times 10^{-6}. 
\end{equation}
Moreover, from the known bounds due to resonator experiments (above),
we know that the components of $\vec{n}$ are of order $10^{-8}$.  This
therefore implies that in the case $\beta < 0$, we must have $|n^0| <
4.9 \times 10^{-6}$.

For the case of photon decay, Hohensee \textit{et al.} note that a
significant fraction of predicted high-energy photons ($E_\gamma
\gtrsim 300$ GeV) have been observed in the D0 detector at Fermilab
(specifically, in the study of isolated-photon production with an
associated jet.)  This implies that the threshold energy for photon
decay cannot greatly exceed 300 GeV; again using the bounds on the
components of $\vec{n}$ from resonator experiments, we find that in
the case $\beta > 0$ we must have
\begin{equation}
  |n^0| \lesssim \frac{2 m_e}{E_\gamma} \approx 3.4 \times 10^{-6}.
\end{equation}

More recently, work by Altschul \cite{AltschulSynch} has extended
these bounds by examining the amount of synchrotron radiation observed
at LEP.  The argument proceeds similarly to the analysis of vacuum
\v{C}erenkov radiation in LEP, given above.  In a Lorentz-violating
theory, the velocity of a relativistic charged particle moving in a
uniform magnetic field will deviate from the ``expected'' velocity
(i.e., that in the absence of Lorentz violation) by $\delta v = \tk$,
where $\tk$ is a linear combination (with coefficients of order unity)
of the Lorentz-violating coefficients $(\tk_{e-})^{ij}$, $(\tk_{o+})$,
and $\tk_\text{tr}$ \cite{AltschulGenSynch}.  The fractional deviation
of the synchrotron power radiated from its expected value is then
given by of order
\begin{equation}
  \frac{\Delta P}{P} = 8 \gamma^2 \tk,
\end{equation}
where $\gamma$ is the boost factor of the relativistic particle.
When we integrate this power deviation over a full cycle of the
charged particle, the parity-odd coefficients must drop out due to
symmetry;  thus, only $(\tk_{e\pm})^{ij}$ and $\tk_\text{tr}$ can in
principle contribute to the integrated synchrotron power loss.

The greatest boost factor $\gamma$ of particles achieved at LEP was
above $2 \times 10^5$; the integrated power loss of particles in the
storage rings at LEP was measured to agree with the predictions of
standard (non-Lorentz-violating) electrodynamics to within a
fractional precision of $|\Delta P|/P < 2 \times 10^{-4}$
\cite{AltschulSynch}.  We can thus conclude that
\begin{equation}
  |\tk| < \frac{1}{8 \gamma^2} \frac{|\Delta P |}{P} \approx 6 \times
  10^{-16}
\end{equation}
Since resonator experiments bound the components of $(\tk_{e-})^{ij}$
to an order of magnitude below this, this bound is therefore only a
bound on $\tk_\text{tr}$.  In our case, $\tk_\text{tr}$ is given by
equation \eqref{kaptrdef}.  We can thus conclude that the known bounds
on synchrotron radiation at LEP limit only the value of
$\tk_\text{tr}$;  specifically,
\begin{equation}
  |n^0| < \sqrt{2 |\tk_\text{tr}|} < 3.5 \times 10^{-8}.
\end{equation}
Note that this is a two-sided bound: since LEP is sensitive to $\Delta
P/P$ being either positive or negative, and since the sign of $\Delta
P/P$ is dependent on $\tk_\text{tr}$, then LEP measurements bound
$\tk_\text{tr}$ both above and below zero.  In the current model, this
means that $n^0$ is bounded both for $\beta >0$ and $\beta <0$.

This bound is comparable to the current bounds on the spatial
components of $\vec{n}$ obtainable from resonator experiments.
However, this bound should be taken correct only to within an order of
magnitude, due to our lack of knowledge about the precise functional
form of $\tk$ in our theory.  A more precise estimate would involve a
full calculation of the rate of synchrotron radiation in our model;
however, such a calculation is outside the scope of this paper.

\subsection{Cosmic ray observations}
\label{CosBdSec}

While man-made particle accelerators can impart very high energies
to particles, it is well-known that ``natural particle accelerators''
elsewhere in the Universe put these efforts to shame;  cosmic rays
have been observed with energies of up to eight orders of magnitude
more energy than the most energetic particles created in the
laboratory to date.  Since our bounds on the parameters of our theory
scale inversely with the energies of observed particles, we will find
that subject to some caveats (detailed below) cosmic ray observations
give us the best bounds on the components of $n^\mu$.

In the case of vacuum \v{C}erenkov radiation, the Pierre Auger
Collaboration has observed a few dozen cosmic ray showers with total
energies above 57 EeV \cite{augerevents}.  It is generally assumed
that the primary particles for such cosmic ray showers are
hyperrelativistic protons or nuclei.  In our case, the inequality
\eqref{Cerineq} becomes more restrictive as the mass of the primary
particle decreases.  Thus, if we wish to place conservative bounds on
the parameters of the theory, we should assume that the primary is a
fairly heavy nucleus.  It is unlikely, however, that these primaries
are significantly heavier than ${}^{56} \text{Fe}$.  We therefore
assume that all the events listed in \cite{augerevents} have a primary
mass of $M_p c_p^2 = 52.1$ GeV.  If these events were to later be
discovered to have a smaller mass $M'_p$, it would simply rescale our
limits on $n^\mu$ by a factor of $M'_p/M_p$.

\begin{table}
  \begin{tabular}{c | r@{.}l @{\hspace{1ex}}  r@{.}l }
    &\multicolumn{2}{c}{Cosmic rays} &
    \multicolumn{2}{c}{VHE photons} 
    \\ \hline \hline
    $|n^0|$ & \hspace{1em} $6$ & $8 \times 10^{-10}$ & \hspace{1em}
    $2$&$1 \times 10^{-8}$  \\
    $|n^X|$ & $5$ & $6 \times 10^{-10}$ & $6$&$7 \times 10^{-8}$ \\
    $|n^Y|$ & $8$ & $2 \times 10^{-10}$ & $3$&$3 \times 10^{-8}$ \\
    $|n^Z|$ & $14$& $6 \times 10^{-10}$ & $5$&$3 \times 10^{-8}$ \\
    \hline 
    & $19$ & $9 \times 10^{-19}$ & $46$&$6 \times 10^{-16}$ \\
    \raisebox{2.5ex}[0pt]{$n^\mu n_\mu$} & $-2$&$3 \times 10^{-19}$
    & $-3$ & $3 \times 10^{-16}$ \\ \hline
    $V_n$ & $6$ & $0 \times 10^{-37}$ & $2$&$7 \times 10^{-30}$ 
  \end{tabular}
  \caption{ \label{cosmictable} Bounds placed on the components of $n^\mu \equiv
    \sqrt{|\beta|} \bb^a$ in the Sun-centred frame by cosmic-ray
    observations ($\beta < 0$) and 
    high-energy gamma-ray observations ($\beta > 0$).}
\end{table}

The 27 events listed in \cite{augerevents} thus disallow large volumes
of parameter space.  The remaining, allowed region of parameter space
is a complicated non-uniform polychoron, symmetric with respect to
reflection about the origin.  The maximum allowed magnitudes of each
of the components of $n^\mu$ are listed in Table \ref{cosmictable},
along with the maximum and minimum allowed values of the norm of
$n^a$, and the volume $V_n$ of the allowed polychoron.

In the case $\beta > 0$, we would expect to see a (possibly
direction-dependent) cutoff in the very high-energy (VHE) gamma-ray
spectrum.  Over the past decade, gamma-ray observatories such as HESS,
VERITAS and MAGIC have catalogued several dozen VHE gamma-ray sources,
and in many cases have been able to associate these sources with known
objects (either in the Milky Way or extragalactic.)  However, two
problems arise when attempting to use these sources to bound the
parameters of our theory.  The first is simply a matter of orders of
magnitude; the ratio of the energy of these gamma-rays to the electron
mass is two orders of magnitude smaller than the boost factor of the
charged particles that make up cosmic rays.  The highest-energy
photons detected thus far have energies of order 75 TeV, yielding a
bounding factor $2 m_e c_p^2/E_\gamma$ of approximately $10^{-8}$; by
contrast, the inverses of the boost factors for the charged-particle
gamma rays observed by the Pierre Auger Observatory are of order
$10^{-10}$.  Thus, we cannot expect nearly so tight a bound on the
components of $n^\mu$ as we obtained in the case $\beta <0$.  

Second, the vast majority of the sources we can reliably use for this
purpose lie in the plane of the Milky Way.  Most known extragalactic
sources of VHE gamma rays are associated with blazars located outside
of the Local Supercluster, with non-negligible cosmological redshift
($z \gtrsim 0.05$.)  Once cosmological effects become non-negligible,
our assumptions of nearly-flat metric \eqref{hdef} and nearly-constant
vector field \eqref{Adef} can no longer be expected to hold.  Instead,
we would expect that $g^{ab}$ would be approximate a
Friedman-Robertson-Walker solution, and that the value of $B^a$ would
be subject to cosmological evolution.  This would cause significant
deviations from the predictions of Section \ref{partkin}, which were
predicated on $\te^{ab}$ being constant throughout the propagation of
the photon.  For the purposes of this paper, we will therefore limit
ourselves to VHE gamma-ray sources known or suspected to be within the
Local Supercluster.  

\begin{table}
  \begin{tabular}{l @{\hspace{1em}} r@{.}l @{\hspace{1em}} r@{.}l r l
      c} 
    Name & \multicolumn{2}{c}{$RA$ (${}^\text{o}$)} &
    \multicolumn{2}{c}{$\delta$ (${}^\text{o}$)} &
    \multicolumn{2}{c} 
    {$E_\text{max}$ (TeV)} & Ref. \\ \hline
    \hline 
    Crab Nebula & 83&6& 22&01 & \hspace{1.5em}75 && \cite{Aharonian:2004p700} \\
    RX J1713.7-3946 & 258&4& $-39$&76 & 47 && \cite{Aharonian:2007} \\
    Vela X & 128&8& $-45$&60 & 45 && \cite{Aharonian:2006p685} \\
    HESS J1825-137 & 276&5 & $-13$&76 & 40 && \cite{Aharonian:2006p715}
    \\ 
    MSH 15-52 & 228&5 & $-59$&16 & 40 && \cite{Aharonian:2005p717} \\
    Galactic Centre & 266&3 & $-29$&00 & 32 && \cite{Aharonian:2009p729}
    \\ 
    HESS J1809-193 & 272&6 & $-19$&30 & 30 && \cite{Aharonian:2007p712}
    \\ 
    HESS J1708-443 & 257&0& $-44$&35 & 20 && \cite{Hoppe:2009p706} \\
    LS 5039 & 276&6& $-14$&85 & 20 && \cite{Aharonian:2006p707} \\
    RCW 86 & 220&7& $-62$&45 & 20 && \cite{Aharonian:2009p709} \\
    HESS J1616-508 & 244&1 & $-50$&90 & 20 && \cite{Aharonian:2006p715}
    \\ 
    HESS J1813-178 & 273&4 & $-17$&84 & 20 && \cite{Aharonian:2006p715}
    \\ 
    M87 & 187&7 & 12&39 & 20 && \cite{Aharonian:2006p738} \\ 
    Westerlund 2 & 155&8 & $-57$&76 & 18 && \cite{Aharonian:2007p728} \\
    HESS J1837-069 & 279&4 & $-6$&95 & 15 && \cite{Aharonian:2006p715}
    \\ 
    Kookaburra & 214&5 & $-60$&98 & 15 && \cite{Aharonian:2006p725} \\
    HESS J1718-385 & 259&5 & $-38$&55 & 15 && \cite{Aharonian:2007p712}
    \\ 
    RX J0852.0-4622 & 133&0 & $-46$&37 & 10 &&
    \cite{Aharonian:2005p643} \\ 
    Cassiopeia A & 350&9 & 58&82 & 10 && \cite{Albert:2007p710} \\
    HESS J1702-420 & 255&7 & $-42$&02 & 10 && \cite{Aharonian:2006p715}
    \\ 
    HESS J1804-216 & 271&1 & $-21$&70 & 10 && \cite{Aharonian:2006p715}
    \\ 
    CTB 37A & 258&6 & $-38$&57 & 10 && \cite{Aharonian:2008p733} \\
    \hline
    Centaurus A & 201&4 & 43&02& 5 && \cite{Aharonian:2009p734} \\
    AE Aquarii & 310&0 & $-0$&87 & 3&\hspace{-0.9ex}.5&
    \cite{Meintjes:1992p647} \\ 
  \end{tabular}
  \caption{\label{gsourcetab} VHE gamma-ray sources with significant
    observed luminosity over 10 TeV.  Shown are the names of the
    sources, their right ascension and declination (in degrees), and
    $E_\text{max}$, the
    approximate maximum energy of gamma rays observed in their
    spectra.  The sources AE Aquarii and Centaurus A have
    $E_\text{max} < 10$ TeV but are included due to their higher
    galactic latitude ($-24.42^\text{o}$ and $19.42^\text{o}$,
    respectively);  see text.}
\end{table}
    
With these caveats in mind, we proceed.  A survey of the literature
reveals 22 known VHE gamma-ray sources with significant flux at
energies greater than 10 TeV (see Table
\ref{gsourcetab}.)\footnote{The large majority of the energies listed
  from these tables are extracted from plots in the references; they
  are therefore necessarily somewhat imprecise.}  Only one of these
sources, M87 \cite{Aharonian:2006p738}, is extragalactic (with $b =
74.49^\text{o}$).  To improve our bounds in the directions orthogonal
to the galactic plane, we therefore include two lower-energy sources:
the radio galaxy Centaurus A \cite{Aharonian:2009p734}, with galactic
latitude $b = -24.42^\text{o}$, and the cataclysmic variable AE
Aquarii \cite{Meintjes:1992p647}, with $b = 19.42^\text{o}$.  The
resulting bounds are shown in Table \ref{cosmictable}.  As expected,
they are significantly less stringent than the bounds imposed by
charged-particle cosmic ray observations; however, they are still
competitive with the limits imposed on the components of $\vec{n}$ by
laboratory and accelerator experiments, and are more stringent in the
case of $n^0$.

\section{Discussion}

Laboratory experiments (both electromagnetic resonators and bounds
from accelerator physics) limit the components of $n^\mu$ to the
$10^{-8}$ level.  In the case $\beta < 0$, cosmic-ray observations
push these bounds down $\mathcal{O}(10^{-10})$ for all components of
$n^\mu$; for $\beta >0$, similar observations bound all the components
of $n^\mu$ to the $10^{-8}$ level.

A few notes concerning the above bounds are in order.  First, the
bounds extracted from the regression in Section \ref{resonsec} are not
based on the current strongest bounds on Lorentz violation in the
photon sector, but rather on older work with more detailed reporting
of data.  One would expect that if the data underlying the most
sensitive experiments to date \cite{HerrCPT,ENS} were analyzed in
this way, our sensitivity to the components of $n^\mu$ would increase
by approximately a factor of three (i.e., half an order of magnitude.)

The great majority of the cosmic-ray sources and events used in
Section \ref{CosBdSec} are in the Southern Celestial Hemisphere.  This
is simply due to the locations of the observatories in question.  In
the case of high-energy charged particles, the Pierre Auger
Observatory (located in Argentina) does not yet have a
Northern-Hemisphere counterpart; Pierre Auger North, a planned
counterpart in Colorado, will not see ``first light'' for some years.
In the case of high-energy gamma rays, this asymmetry is simply due to
the fact that the HESS telescope (located in Namibia) has been in
operation longer than the similar MAGIC and VERITAS telescopes
(located in the Canary Islands and Arizona, respectively.)  As these
latter two telescopes and the air-shower arrays MILAGRO (New Mexico),
AS$\gamma$ (Tibet), and ARGO (Tibet) report more high-energy sources
(especially in the region of the galactic plane lying in the Northern
Celestial Hemisphere), we can expect these bounds to become more
symmetric.\footnote{Note that a pair of (hypothetical) antipodal
  sources do not place the same bounds on the components of $n^\mu$;
  one bounds $|n^0 + \vec{n} \cdot \hk|$, while the other bounds $|n^0
  - \vec{n} \cdot \hk|$.}

Finally, it is important to note that there are fundamental limits on
how well gamma-ray experiments can bound the parameters of our
theory. Photons with sufficiently high energies can interact with
background radiation (particularly the cosmic microwave background)
and produce electron-positron pairs \cite{GRH}.  In particular, the
``gamma-ray horizon'' for $\sim 100$ TeV gamma rays is approximately
the size of the Local Supercluster, and for $\sim 1$ PeV gamma rays is
approximately the size of the Milky Way.  This pair-production
attenuation therefore places fundamental bounds on the region of
parameter space that can be bounded by gamma-ray observations: we do
not expect to get extremely high-energy gamma rays from outside our
own galaxy, but our bounds on the component of $\vec{n}$ orthogonal to
the galactic plane will generically be insensitive to sources from
within the Galaxy.

\acknowledgments

I am indebted to V.~A.~Kosteleck\'{y}, S.~Parker, P.~Stanwix, and
M.~Tobar for their helpful discussions and correspondence.  The
comments of the anonymous referee are also gratefully acknowledged.
The TeVCat website (\texttt{http://tevcat.uchicago.edu}), maintained
by S.~Wakely and D.~Horan, was indispensable in the preparation of
Table \ref{gsourcetab}.  This work was supported in part by the United
States Department of Energy, under grant DE-FG02-91ER40661.

\appendix*

\section{Photon Lifetime}

In using particle physics to place bounds on Lorentz-violating
theories, it is informative to examine the mean free paths of
particles or, equivalently, their lifetimes.  The rate of energy loss
to vacuum \v{C}erenkov radiation in a general Lorentz-violating theory
has been previously derived \cite{AltschulCer}.  However, the decay
$\gamma \to e^+ e^-$, as occurs in our case when $\beta > 0$, is less
well-understood.  The full quantum theory of Lorentz-violating photons
is not yet known, and so a calculation of the exact photon decay rate
is impossible.  However, it is well-known that when certain
Lorentz-violating coefficients are sufficiently small, they can be
``moved'' between sectors via redefinitions of the metric (or,
equivalently, redefinitions of the coordinates.) In particular, for a
theory of Lorentz-violating photons without vacuum birefringence on a
flat metric, it is possible to shift the Lorentz-violating
coefficients entirely into the electron sector instead \cite{LehKos}.
The quantum field theory of Lorentz-violating electrons being much
better understood \cite{CKcross}, we pursue this tactic.  We will see
that such a technique will only yield decay rates that are accurate to
first order in $\beta$; however, this calculation will still be
valuable in estimating the mean free paths of Lorentz-violating
photons.

A similar calculation for ``isotropic Lorentz-violating photons'' has
been performed by Hohensee \textit{et al.}~\cite{Hohensee};  we will
follow their techniques here.  For consistency with this paper and
\cite{CKcross}, we use a metric with signature $(+,-,-,-)$.  In
particular, when switching to this sign convention, our definition of
$\te^{ab}$ will change;  in this sign convention, a metric of the form
\begin{equation}
  \label{newtedef}
  \te^{ab} = \eta^{ab} - \beta \bb^a \bb^b
\end{equation}
is physically equivalent to the metric used in the previous sections.

We start with the ``Lorentz-violating QED'' Lagrangian for bumblebee
photons minimally coupled to an electron field $\psi$ in flat spacetime:
\begin{equation}
  \label{LVQED}
  \mathcal{L} = \frac{i}{2} \bar{\psi} \gamma^a
  \doubarrow{D}_a \psi - m \bar{\psi} \psi -
  \frac{1}{4} \te^{ab} \te^{cd} F_{ac} F_{bd},
\end{equation}
with $\te^{ab}$ defined as in \eqref{newtedef}.  The process used by
Hohensee \textit{et al.}~to derive the photon lifetime consists of
several steps:
\begin{enumerate}
\item We redefine the metric so that the ``photon metric'' $\te^{ab}$
  is the ``true metric'' of the theory.  Since the gamma matrices in
  the electron kinetic term are defined with respect to the original
  metric $\eta^{ab}$ (i.e., $\{ \gamma^a, \gamma^b \} = 2 \eta^{ab}$),
  we must rewrite these matrices in terms of new gamma matrices
  $\tilde{\gamma}^a$ defined such that $\{ \tilde{\gamma}^a,
  \tilde{\gamma}^b \} = 2 \te^{ab}$.  To first order in $\beta$, these
  matrices are related by
  \begin{equation}
    \label{newgammadef}
    \gamma^a = \tilde{\gamma}^a + \frac{\beta}{2} \bb^a \bb^b \te_{bc}
    \tilde{\gamma}^c. 
  \end{equation}
  Applying this definition to our Lagrangian, we find that our new
  Lagrangian is
  \begin{equation}
    \mathcal{L} = \frac{i}{2} \bar{\psi} \left( \tilde{\gamma}^a +
      \frac{\beta}{2} \bb^a \bb^b \tilde{\gamma}_b \right)
    \doubarrow{D}_a \psi - m \bar{\psi} \psi - \frac{1}{4} F_{ab} F^{ab}
  \end{equation}
  where indices are now raised and lowered with the metric
  $\te^{ab}$.  We will hereafter ``drop the tildes'' for notational
  convenience.  

\item The kinetic term for the electron now contains non-standard time
  derivatives; these must be eliminated to successfully quantize the
  theory \cite{LehKos}.  To do this, we define a new spinor field
  $\chi$ such that $\psi = A \chi$, where the matrix $A$ satisfies
  \begin{equation}
    A^\dag \gamma^0 \left( \gamma^0 +
      \frac{\beta}{2} \bb^0 \bb^b \gamma_b \right) A =
    \mathbf{1}.
  \end{equation}
  To first order in $\beta$, this implies that
  \begin{equation}
    A = \mathbf{1} - \frac{\beta}{4} \bb^0 \bb_a \gamma^0 \gamma^a.
  \end{equation}
  Rewriting the Lagrangian in terms of $\chi$ then yields
  \begin{equation}
    \mathcal{L} = \frac{i}{2} \he^{ab} \bar{\chi} \gamma_a
    \doubarrow{D}_b \chi - \hat{m} \bar{\chi} \chi - \frac{1}{4}
    F_{ab} F^{ab},
  \end{equation}
  where
  \begin{equation}
    \he^{ab} = \eta^{ab} - \frac{\beta}{2} \left( (\bb^0)^2
      \eta^{ab} + 2 \bb^0 \bb^{[a} \eta^{b]0} - \bb^a \bb^b \right)
  \end{equation}
  and
  \begin{equation}
    \hat{m} = m \left( 1 - \frac{\beta}{2} (\bb^0)^2 \right).
  \end{equation} 

\item Write down the invariant matrix element $\mathcal{M}$.
  Using the photon and electron polarizations defined in
  \cite{Hohensee}, we find this to be given by
  \begin{equation}
    i \mathcal{M}_{rs} = - i e \epsilon_a (p) \he^{ab}
    \bar{u}^{(r)} (q) \gamma_b v^{(s)} (k)
  \end{equation}
  where $p^a$ is the incoming photon momentum, $\epsilon_a(p)$ is its
  polarization, $q^a$ and $k^a$ are the outgoing electron and positron
  momenta respectively, and $\bar{u}^{(r)} (q)$ and $v^{(s)} (k)$ are their
  respective polarization states.  To obtain the photon lifetime, we
  will need to sum over the final fermion polarization states and
  average over photon polarization states;  using the trace identities
  derived in \cite{CKcross}, this yields
  \begin{align}
    \overline{| \mathcal{M}|^2} &= \frac{1}{2} \sum_\epsilon
    \sum_{r,s} |\mathcal{M}_{r,s} |^2 \notag \\
    &= 2 e^2 \big( \he^{ab} \he_{ab} (\hat{m}^2 +
      \he^e {}_c \he_{ed} q^c k^d ) \notag  \\ & \mspace{150mu} {} - 2 \he_a {}^b
        \he_{cb} \he^{ad} \he^{ce} q_d k_e
      \big).
      \label{matrixelem}
  \end{align}
  We have used the spinor normalization conventions chosen in
  \cite{Hohensee} (i.e., $N(\vec{q}) = 2 E_q$ and similarly for $k$.)
  
\item Integrate over the final electron and positron momenta to
  obtain the photon lifetime as a function of its energy $E_\gamma$ and
  propagation direction $\hat{p}$:
  \begin{equation}
    \Gamma(\hat{p}) = \frac{1}{4 \pi^2} \frac{1}{2 E_\gamma} \int \frac{ \dif^3
      \vec{q}}{2 E_q} \frac{\dif^3 \vec{k}}{2 E_k}
    \overline{|\mathcal{M}|^2} \delta^4 (p^\mu - q^\mu - k^\mu)
  \end{equation}

\end{enumerate}

I have as yet been unable to derive a closed-form analytical
expression for this lifetime.  However, an order-of-magnitude estimate
of the photon decay rate may be obtained by estimating the quantity
$\overline{|\mathcal{M}|^2} / 4 E_q E_k$ for an on-shell decay and
multiplying this quantity by the allowed volume of phase space; in
other words,
\begin{multline}
  \int \frac{ \dif^3
    \vec{q}}{2 E_q} \frac{\dif^3 \vec{k}}{2 E_k}
  \overline{|\mathcal{M}|^2} \delta^4 (p^\mu - q^\mu - k^\mu) \notag
  \\ =
  \left\langle \frac{\overline{|\mathcal{M}|^2}}{4 E_q E_k}
  \right\rangle \times \int \dif^3 \vec{q} \, \dif^3 \vec{k} \, \delta^4
  (p^\mu - q^\mu - k^\mu) 
  \label{lifetimedecomp}
\end{multline}
This equation can be thought of as defining the quantity in angle
brackets above.  To estimate its order of magnitude, we can use the
dispersion relation for the electrons in this theory,
\begin{equation}
  \he_{ac} \he_b {}^c k^a k^b = \hat{m}^2,
\end{equation}
along with momentum conservation, $p^a = k^a + q^a$, to put
\eqref{matrixelem} in the form
\begin{multline}
  \overline{|\mathcal{M}|^2} = 2 e^2 \Big[
    2 \beta E_\gamma^2 \left( - 2 \bb^0 \vec{B} \cdot \hat{p} +
      (\vec{B} \cdot \hat{p})^2 
    \right) \\ + 2 (1 - \beta (\bb^0)^2) \hat{m}^2 + \beta  (
      \bb_a k^a)^2 + \beta (
      \bb_a q^a)^2  \Big] \\ + \mathcal{O}(\beta^2)
      \label{matelen}
\end{multline}
All of these terms except for the last two are constant over the mass
shell.  To estimate the order of magnitude of these last two terms, we
note that the electron dispersion relation satisfies $E_k^2 -
\vec{k}^2 = m^2 + \mathcal{O}(\beta)$;  thus,
\begin{align}
  \beta (\bb_a k^a)^2 &= \beta (\bb^0 E_k - |\vec{B}| |\vec{k}| \cos
  \theta )^2 \notag \\
  &\sim \beta \left(\bb^0 E_k - |\vec{B}| \sqrt{E_k^2 - m^2} \right)^2
\end{align}
where we have discarded the $\cos \theta$ term because it is of order
unity.  We can further estimate that for a generic decay $E_k$ (and
$E_q$) will be of order $E_\gamma/2$, and that (since $E_\gamma >
E_{d0} \gg m^2$) the resulting electrons will have relativistic
velocities.  This then implies that
\begin{equation}
  \beta (\bb_a k^a)^2 \sim \beta (\bb_a q^a)^2 \sim \beta
  \mathcal{B}^2 \frac{E_\gamma^2}{4}   
\end{equation}
where $\mathcal{B}$ is a quantity of the same order as the components
of $\bb^a$.  We can then estimate the value of
$\overline{|\mathcal{M}|^2}$ to be
\begin{equation}
\overline{|\mathcal{M}|^2} \sim 2 e^2 \left[ \beta \mathcal{B}^2
  E_\gamma^2 + 2 (1 - \beta (\bb^0)^2) \hat{m}^2 \right] \sim 2 e^2
\beta \mathcal{B}^2 E_\gamma^2,
\end{equation}
where we have redefined $\mathcal{B}$ to include the contributions of
the first term in \eqref{matelen}.  (Note that the second term in the
equation above is of equal or lesser magnitude than the first, since
$\beta \mathcal{B}^2 E_\gamma^2 > \beta \mathcal{B}^2 E_{d0}^2 \sim
m^2$.) 

 Thus, we can estimate that
\begin{equation}
\left\langle \frac{\overline{|\mathcal{M}|^2}}{4 E_q E_k}
  \right\rangle \sim 8 \pi \alpha \beta \mathcal{B}^2,
\end{equation}
where $\alpha$ is the fine structure constant.  The volume of
kinematically accessible phase space, meanwhile, can be shown to be
\begin{multline}
\int \dif^3 \vec{q} \, \dif^3 \vec{k} \, \delta^4
  (p^\mu - q^\mu - k^\mu) \\ = \frac{\pi}{2} E_\gamma^2 \sqrt{1 -
    \frac{E_{d0}^2}{E_\gamma^2}} + \mathcal{O}(\beta) + \mathcal{O}
  \left( \frac{m^2}{E_\gamma^2} \right)
\end{multline}
This integral is most easily done by changing coordinates on phase
space to $\vec{u} = \vec{q} + \vec{k}$ and $\vec{v} =
\frac{1}{2}(\vec{q} - \vec{k})$; the resulting expression sets
$\vec{u} = \vec{p}$, leaving an expression proportional to the volume
of an ellipsoidal shell in $\vec{v}$-space.  We can then perform a
further linear transformation on $\vec{v}$ to make this shell
spherical.  Combining these results, we can then estimate that the
photon lifetime is 
\begin{equation}
  \Gamma \sim \frac{\alpha}{2} \beta \mathcal{B}^2 E_\gamma \sqrt{1 -
    \frac{E_{d0}^2}{E_\gamma^2}} 
\end{equation}
to within an order of magnitude.  Note that the scaling behaviour of
this result is consistent with the exact results derived in the
spatially isotropic case by Hohensee \emph{et al.} \cite{Hohensee}.

\bibliographystyle{apsrev}
\bibliography{gbm}{}

\begin{thebibliography}{45}
\expandafter\ifx\csname natexlab\endcsname\relax\def\natexlab#1{#1}\fi
\expandafter\ifx\csname bibnamefont\endcsname\relax
  \def\bibnamefont#1{#1}\fi
\expandafter\ifx\csname bibfnamefont\endcsname\relax
  \def\bibfnamefont#1{#1}\fi
\expandafter\ifx\csname citenamefont\endcsname\relax
  \def\citenamefont#1{#1}\fi
\expandafter\ifx\csname url\endcsname\relax
  \def\url#1{\texttt{#1}}\fi
\expandafter\ifx\csname urlprefix\endcsname\relax\def\urlprefix{URL }\fi
\providecommand{\bibinfo}[2]{#2}
\providecommand{\eprint}[2][]{\url{#2}}

\bibitem[{\citenamefont{Kosteleck{\'y} and Russell}()}]{LVdata}
\bibinfo{author}{\bibfnamefont{V.~A.} \bibnamefont{Kosteleck{\'y}}}
  \bibnamefont{and} \bibinfo{author}{\bibfnamefont{N.}~\bibnamefont{Russell}},
  \bibinfo{note}{ar$\chi$iv:0801.0287v2}.

\bibitem[{\citenamefont{Colladay and Kosteleck{\'y}}(1997)}]{CPTviol}
\bibinfo{author}{\bibfnamefont{D.}~\bibnamefont{Colladay}} \bibnamefont{and}
  \bibinfo{author}{\bibfnamefont{V.~A.} \bibnamefont{Kosteleck{\'y}}},
  \bibinfo{journal}{Phys. Rev. D} \textbf{\bibinfo{volume}{55}},
  \bibinfo{pages}{6760} (\bibinfo{year}{1997}).

\bibitem[{\citenamefont{Colladay and Kosteleck{\'y}}(1998)}]{CKSME}
\bibinfo{author}{\bibfnamefont{D.}~\bibnamefont{Colladay}} \bibnamefont{and}
  \bibinfo{author}{\bibfnamefont{V.~A.} \bibnamefont{Kosteleck{\'y}}},
  \bibinfo{journal}{Phys. Rev. D} \textbf{\bibinfo{volume}{58}},
  \bibinfo{pages}{116002} (\bibinfo{year}{1998}).

\bibitem[{\citenamefont{Kosteleck{\'y}}(2004)}]{Kosgrav}
\bibinfo{author}{\bibfnamefont{V.~A.} \bibnamefont{Kosteleck{\'y}}},
  \bibinfo{journal}{Phys. Rev. D} \textbf{\bibinfo{volume}{69}},
  \bibinfo{pages}{105009} (\bibinfo{year}{2004}).

\bibitem[{\citenamefont{Seifert}(2009)}]{VMLB}
\bibinfo{author}{\bibfnamefont{M.~D.} \bibnamefont{Seifert}},
  \bibinfo{journal}{Phys. Rev. D} \textbf{\bibinfo{volume}{79}},
  \bibinfo{pages}{124012} (\bibinfo{year}{2009}).

\bibitem[{\citenamefont{Bluhm and Kosteleck{\'y}}(2005)}]{BKNG}
\bibinfo{author}{\bibfnamefont{R.}~\bibnamefont{Bluhm}} \bibnamefont{and}
  \bibinfo{author}{\bibfnamefont{V.~A.} \bibnamefont{Kosteleck{\'y}}},
  \bibinfo{journal}{Phys. Rev. D} \textbf{\bibinfo{volume}{71}},
  \bibinfo{pages}{065008} (\bibinfo{year}{2005}).

\bibitem[{\citenamefont{Bluhm et~al.}(2008)\citenamefont{Bluhm, Fung, and
  Kosteleck{\'y}}}]{BFKmass}
\bibinfo{author}{\bibfnamefont{R.}~\bibnamefont{Bluhm}},
  \bibinfo{author}{\bibfnamefont{S.-H.} \bibnamefont{Fung}}, \bibnamefont{and}
  \bibinfo{author}{\bibfnamefont{V.~A.} \bibnamefont{Kosteleck{\'y}}},
  \bibinfo{journal}{Phys. Rev. D} \textbf{\bibinfo{volume}{77}},
  \bibinfo{pages}{065020} (\bibinfo{year}{2008}).

\bibitem[{\citenamefont{Wald}(1984)}]{Wald}
\bibinfo{author}{\bibfnamefont{R.~M.} \bibnamefont{Wald}},
  \emph{\bibinfo{title}{General Relativity}} (\bibinfo{publisher}{University of
  Chicago Press}, \bibinfo{year}{1984}).

\bibitem[{\citenamefont{Isenberg and Nester}(1977)}]{IsenNest}
\bibinfo{author}{\bibfnamefont{J.}~\bibnamefont{Isenberg}} \bibnamefont{and}
  \bibinfo{author}{\bibfnamefont{J.}~\bibnamefont{Nester}},
  \bibinfo{journal}{Ann. Phys.} \textbf{\bibinfo{volume}{107}},
  \bibinfo{pages}{56} (\bibinfo{year}{1977}).

\bibitem[{\citenamefont{Bailey and Kosteleck\'y}(2006)}]{BK}
\bibinfo{author}{\bibfnamefont{Q.~G.} \bibnamefont{Bailey}} \bibnamefont{and}
  \bibinfo{author}{\bibfnamefont{V.~A.} \bibnamefont{Kosteleck\'y}},
  \bibinfo{journal}{Phys. Rev. D} \textbf{\bibinfo{volume}{74}},
  \bibinfo{pages}{045001} (\bibinfo{year}{2006}).

\bibitem[{\citenamefont{Jacobson and Mattingly}(2001)}]{einaeth}
\bibinfo{author}{\bibfnamefont{T.}~\bibnamefont{Jacobson}} \bibnamefont{and}
  \bibinfo{author}{\bibfnamefont{D.}~\bibnamefont{Mattingly}},
  \bibinfo{journal}{Phys. Rev. D} \textbf{\bibinfo{volume}{64}},
  \bibinfo{pages}{024028} (\bibinfo{year}{2001}).

\bibitem[{\citenamefont{Kosteleck\'y and Samuel}(1989)}]{KosSam}
\bibinfo{author}{\bibfnamefont{V.~A.} \bibnamefont{Kosteleck\'y}}
  \bibnamefont{and} \bibinfo{author}{\bibfnamefont{S.}~\bibnamefont{Samuel}},
  \bibinfo{journal}{Phys. Rev. D} \textbf{\bibinfo{volume}{40}},
  \bibinfo{pages}{1886} (\bibinfo{year}{1989}).

\bibitem[{\citenamefont{Arkani-Hamed et~al.}(2005)\citenamefont{Arkani-Hamed,
  Cheng, Luty, and Thaler}}]{AHspinforce}
\bibinfo{author}{\bibfnamefont{N.}~\bibnamefont{Arkani-Hamed}},
  \bibinfo{author}{\bibfnamefont{H.-C.} \bibnamefont{Cheng}},
  \bibinfo{author}{\bibfnamefont{M.}~\bibnamefont{Luty}}, \bibnamefont{and}
  \bibinfo{author}{\bibfnamefont{J.}~\bibnamefont{Thaler}},
  \bibinfo{journal}{J. High Energy Phys.} \textbf{\bibinfo{volume}{07(2005)}},
  \bibinfo{pages}{029} (\bibinfo{year}{2005}).

\bibitem[{\citenamefont{Kosteleck{\'y} and Tasson}(2009)}]{KTgrava}
\bibinfo{author}{\bibfnamefont{V.~A.} \bibnamefont{Kosteleck{\'y}}}
  \bibnamefont{and} \bibinfo{author}{\bibfnamefont{J.~D.}
  \bibnamefont{Tasson}}, \bibinfo{journal}{Phys. Rev. Lett.}
  \textbf{\bibinfo{volume}{102}}, \bibinfo{pages}{010402}
  (\bibinfo{year}{2009}).

\bibitem[{\citenamefont{Kosteleck{\'y} and Lehnert}(2001)}]{LehKos}
\bibinfo{author}{\bibfnamefont{V.~A.} \bibnamefont{Kosteleck{\'y}}}
  \bibnamefont{and} \bibinfo{author}{\bibfnamefont{R.}~\bibnamefont{Lehnert}},
  \bibinfo{journal}{Phys. Rev. D} \textbf{\bibinfo{volume}{63}},
  \bibinfo{pages}{065008} (\bibinfo{year}{2001}).

\bibitem[{\citenamefont{Kosteleck{\'y} and Mewes}(2002)}]{KMsig}
\bibinfo{author}{\bibfnamefont{V.~A.} \bibnamefont{Kosteleck{\'y}}}
  \bibnamefont{and} \bibinfo{author}{\bibfnamefont{M.}~\bibnamefont{Mewes}},
  \bibinfo{journal}{Phys. Rev. D} \textbf{\bibinfo{volume}{66}},
  \bibinfo{pages}{056005} (\bibinfo{year}{2002}).

\bibitem[{\citenamefont{Lehnert and Potting}(2004)}]{ChernSimonsphoton}
\bibinfo{author}{\bibfnamefont{R.}~\bibnamefont{Lehnert}} \bibnamefont{and}
  \bibinfo{author}{\bibfnamefont{R.}~\bibnamefont{Potting}},
  \bibinfo{journal}{Phys. Rev. Lett.} \textbf{\bibinfo{volume}{93}},
  \bibinfo{pages}{110402} (\bibinfo{year}{2004}).

\bibitem[{\citenamefont{Altschul}(2008)}]{AltschulCer}
\bibinfo{author}{\bibfnamefont{B.}~\bibnamefont{Altschul}},
  \bibinfo{journal}{Nucl.~Phys. B} \textbf{\bibinfo{volume}{796}},
  \bibinfo{pages}{262} (\bibinfo{year}{2008}).

\bibitem[{\citenamefont{Herrmann et~al.}(2008)\citenamefont{Herrmann, Senger,
  M\"ohle, Kovalchuk, and Peters}}]{HerrCPT}
\bibinfo{author}{\bibfnamefont{S.}~\bibnamefont{Herrmann}},
  \bibinfo{author}{\bibfnamefont{A.}~\bibnamefont{Senger}},
  \bibinfo{author}{\bibfnamefont{K.}~\bibnamefont{M\"ohle}},
  \bibinfo{author}{\bibfnamefont{E.~V.} \bibnamefont{Kovalchuk}},
  \bibnamefont{and} \bibinfo{author}{\bibfnamefont{A.}~\bibnamefont{Peters}},
  in \emph{\bibinfo{booktitle}{Proceedings of the Fourth Meeting on CPT and
  Lorentz Symmetry}} (\bibinfo{year}{2008}), pp. \bibinfo{pages}{9--15}.

\bibitem[{\citenamefont{Eisele et~al.}(2009)\citenamefont{Eisele, \relax{Yu}.
  Nevsky, and Schiller}}]{ENS}
\bibinfo{author}{\bibfnamefont{{\relax Ch}.}~\bibnamefont{Eisele}},
  \bibinfo{author}{\bibfnamefont{A.}~\bibnamefont{\relax{Yu}. Nevsky}},
  \bibnamefont{and} \bibinfo{author}{\bibfnamefont{S.}~\bibnamefont{Schiller}},
  \bibinfo{journal}{Phys. Rev. Lett.} \textbf{\bibinfo{volume}{103}},
  \bibinfo{pages}{090401} (\bibinfo{year}{2009}).

\bibitem[{\citenamefont{Stanwix et~al.}(2006)\citenamefont{Stanwix, Tobar,
  Wolf, Locke, and Ivanov}}]{Stan06}
\bibinfo{author}{\bibfnamefont{P.~L.} \bibnamefont{Stanwix}},
  \bibinfo{author}{\bibfnamefont{M.~E.} \bibnamefont{Tobar}},
  \bibinfo{author}{\bibfnamefont{P.}~\bibnamefont{Wolf}},
  \bibinfo{author}{\bibfnamefont{C.~R.} \bibnamefont{Locke}}, \bibnamefont{and}
  \bibinfo{author}{\bibfnamefont{E.~N.} \bibnamefont{Ivanov}},
  \bibinfo{journal}{Phys. Rev. D} \textbf{\bibinfo{volume}{74}},
  \bibinfo{pages}{081101(R)} (\bibinfo{year}{2006}).

\bibitem[{\citenamefont{Hohensee et~al.}(2009)\citenamefont{Hohensee, Lehnert,
  Phillips, and Walsworth}}]{Hohensee}
\bibinfo{author}{\bibfnamefont{M.~A.} \bibnamefont{Hohensee}},
  \bibinfo{author}{\bibfnamefont{R.}~\bibnamefont{Lehnert}},
  \bibinfo{author}{\bibfnamefont{D.~F.} \bibnamefont{Phillips}},
  \bibnamefont{and} \bibinfo{author}{\bibfnamefont{R.~L.}
  \bibnamefont{Walsworth}}, \bibinfo{journal}{Phys. Rev. D}
  \textbf{\bibinfo{volume}{80}}, \bibinfo{pages}{036010}
  (\bibinfo{year}{2009}).

\bibitem[{\citenamefont{Altschul}()}]{AltschulSynch}
\bibinfo{author}{\bibfnamefont{B.}~\bibnamefont{Altschul}},
  \eprint{ar$\chi$iv:0905.4346}.

\bibitem[{\citenamefont{Altschul}(2005)}]{AltschulGenSynch}
\bibinfo{author}{\bibfnamefont{B.}~\bibnamefont{Altschul}},
  \bibinfo{journal}{Phys. Rev. D} \textbf{\bibinfo{volume}{72}},
  \bibinfo{pages}{085003} (\bibinfo{year}{2005}).

\bibitem[{\citenamefont{Abraham et~al.}(2008)\citenamefont{Abraham, Abreu,
  Aglietta, Aguirre, Allard, Allekotte, Allen, Allison, Alvarez-Mu{\~n}iz,
  Ambrosio et~al.}}]{augerevents}
\bibinfo{author}{\bibfnamefont{J.}~\bibnamefont{Abraham}},
  \bibinfo{author}{\bibfnamefont{P.}~\bibnamefont{Abreu}},
  \bibinfo{author}{\bibfnamefont{M.}~\bibnamefont{Aglietta}},
  \bibinfo{author}{\bibfnamefont{C.}~\bibnamefont{Aguirre}},
  \bibinfo{author}{\bibfnamefont{D.}~\bibnamefont{Allard}},
  \bibinfo{author}{\bibfnamefont{I.}~\bibnamefont{Allekotte}},
  \bibinfo{author}{\bibfnamefont{J.}~\bibnamefont{Allen}},
  \bibinfo{author}{\bibfnamefont{P.}~\bibnamefont{Allison}},
  \bibinfo{author}{\bibfnamefont{J.}~\bibnamefont{Alvarez-Mu{\~n}iz}},
  \bibinfo{author}{\bibfnamefont{M.}~\bibnamefont{Ambrosio}},
  \bibnamefont{et~al.}, \bibinfo{journal}{Astropart. Phys.}
  \textbf{\bibinfo{volume}{29}}, \bibinfo{pages}{188} (\bibinfo{year}{2008}).

\bibitem[{\citenamefont{Aharonian et~al.}(2004)\citenamefont{Aharonian,
  Akhperjanian, Beilicke, Bernl{\"o}hr, B{\"o}rst, Bojahr, Bolz, Coarasa,
  Contreras, Cortina et~al.}}]{Aharonian:2004p700}
\bibinfo{author}{\bibfnamefont{F.}~\bibnamefont{Aharonian}},
  \bibinfo{author}{\bibfnamefont{A.}~\bibnamefont{Akhperjanian}},
  \bibinfo{author}{\bibfnamefont{M.}~\bibnamefont{Beilicke}},
  \bibinfo{author}{\bibfnamefont{K.}~\bibnamefont{Bernl{\"o}hr}},
  \bibinfo{author}{\bibfnamefont{H.-G.} \bibnamefont{B{\"o}rst}},
  \bibinfo{author}{\bibfnamefont{H.}~\bibnamefont{Bojahr}},
  \bibinfo{author}{\bibfnamefont{O.}~\bibnamefont{Bolz}},
  \bibinfo{author}{\bibfnamefont{T.}~\bibnamefont{Coarasa}},
  \bibinfo{author}{\bibfnamefont{J.~L.} \bibnamefont{Contreras}},
  \bibinfo{author}{\bibfnamefont{J.}~\bibnamefont{Cortina}},
  \bibnamefont{et~al.}, \bibinfo{journal}{Astrophys. J.}
  \textbf{\bibinfo{volume}{614}}, \bibinfo{pages}{897} (\bibinfo{year}{2004}).

\bibitem[{\citenamefont{Aharonian
  et~al.}(2007{\natexlab{a}})\citenamefont{Aharonian, Akhperjanian,
  Bazer-Bachi, Beilicke, Benbow, Berge, Bernl{\"o}hr, Boisson, Bolz, Borrel
  et~al.}}]{Aharonian:2007}
\bibinfo{author}{\bibfnamefont{F.}~\bibnamefont{Aharonian}},
  \bibinfo{author}{\bibfnamefont{A.~G.} \bibnamefont{Akhperjanian}},
  \bibinfo{author}{\bibfnamefont{A.~R.} \bibnamefont{Bazer-Bachi}},
  \bibinfo{author}{\bibfnamefont{M.}~\bibnamefont{Beilicke}},
  \bibinfo{author}{\bibfnamefont{W.}~\bibnamefont{Benbow}},
  \bibinfo{author}{\bibfnamefont{D.}~\bibnamefont{Berge}},
  \bibinfo{author}{\bibfnamefont{K.}~\bibnamefont{Bernl{\"o}hr}},
  \bibinfo{author}{\bibfnamefont{C.}~\bibnamefont{Boisson}},
  \bibinfo{author}{\bibfnamefont{O.}~\bibnamefont{Bolz}},
  \bibinfo{author}{\bibfnamefont{V.}~\bibnamefont{Borrel}},
  \bibnamefont{et~al.}, \bibinfo{journal}{Astron. Astrophys.}
  \textbf{\bibinfo{volume}{464}}, \bibinfo{pages}{235}
  (\bibinfo{year}{2007}{\natexlab{a}}).

\bibitem[{\citenamefont{Aharonian
  et~al.}(2006{\natexlab{a}})\citenamefont{Aharonian, Akhperjanian,
  Bazer-Bachi, Beilicke, Benbow, Berge, Bernl{\"o}hr, Boisson, Bolz, Borrel
  et~al.}}]{Aharonian:2006p685}
\bibinfo{author}{\bibfnamefont{F.}~\bibnamefont{Aharonian}},
  \bibinfo{author}{\bibfnamefont{A.~G.} \bibnamefont{Akhperjanian}},
  \bibinfo{author}{\bibfnamefont{A.~R.} \bibnamefont{Bazer-Bachi}},
  \bibinfo{author}{\bibfnamefont{M.}~\bibnamefont{Beilicke}},
  \bibinfo{author}{\bibfnamefont{W.}~\bibnamefont{Benbow}},
  \bibinfo{author}{\bibfnamefont{D.}~\bibnamefont{Berge}},
  \bibinfo{author}{\bibfnamefont{K.}~\bibnamefont{Bernl{\"o}hr}},
  \bibinfo{author}{\bibfnamefont{C.}~\bibnamefont{Boisson}},
  \bibinfo{author}{\bibfnamefont{O.}~\bibnamefont{Bolz}},
  \bibinfo{author}{\bibfnamefont{V.}~\bibnamefont{Borrel}},
  \bibnamefont{et~al.}, \bibinfo{journal}{Astron. Astrophys.}
  \textbf{\bibinfo{volume}{448}}, \bibinfo{pages}{L43}
  (\bibinfo{year}{2006}{\natexlab{a}}).

\bibitem[{\citenamefont{Aharonian
  et~al.}(2006{\natexlab{b}})\citenamefont{Aharonian, Akhperjanian,
  Bazer-Bachi, Beilicke, Benbow, Berge, Bernl{\"o}hr, Boisson, Bolz, Borrel
  et~al.}}]{Aharonian:2006p715}
\bibinfo{author}{\bibfnamefont{F.}~\bibnamefont{Aharonian}},
  \bibinfo{author}{\bibfnamefont{A.~G.} \bibnamefont{Akhperjanian}},
  \bibinfo{author}{\bibfnamefont{A.~R.} \bibnamefont{Bazer-Bachi}},
  \bibinfo{author}{\bibfnamefont{M.}~\bibnamefont{Beilicke}},
  \bibinfo{author}{\bibfnamefont{W.}~\bibnamefont{Benbow}},
  \bibinfo{author}{\bibfnamefont{D.}~\bibnamefont{Berge}},
  \bibinfo{author}{\bibfnamefont{K.}~\bibnamefont{Bernl{\"o}hr}},
  \bibinfo{author}{\bibfnamefont{C.}~\bibnamefont{Boisson}},
  \bibinfo{author}{\bibfnamefont{O.}~\bibnamefont{Bolz}},
  \bibinfo{author}{\bibfnamefont{V.}~\bibnamefont{Borrel}},
  \bibnamefont{et~al.}, \bibinfo{journal}{Astrophys. J.}
  \textbf{\bibinfo{volume}{636}}, \bibinfo{pages}{777}
  (\bibinfo{year}{2006}{\natexlab{b}}).

\bibitem[{\citenamefont{Aharonian
  et~al.}(2005{\natexlab{a}})\citenamefont{Aharonian, Akhperjanian, Aye,
  Bazer-Bachi, Beilicke, Benbow, Berge, Berghaus, Bernl{\"o}hr, Boisson
  et~al.}}]{Aharonian:2005p717}
\bibinfo{author}{\bibfnamefont{F.}~\bibnamefont{Aharonian}},
  \bibinfo{author}{\bibfnamefont{A.~G.} \bibnamefont{Akhperjanian}},
  \bibinfo{author}{\bibfnamefont{K.-M.} \bibnamefont{Aye}},
  \bibinfo{author}{\bibfnamefont{A.~R.} \bibnamefont{Bazer-Bachi}},
  \bibinfo{author}{\bibfnamefont{M.}~\bibnamefont{Beilicke}},
  \bibinfo{author}{\bibfnamefont{W.}~\bibnamefont{Benbow}},
  \bibinfo{author}{\bibfnamefont{D.}~\bibnamefont{Berge}},
  \bibinfo{author}{\bibfnamefont{P.}~\bibnamefont{Berghaus}},
  \bibinfo{author}{\bibfnamefont{K.}~\bibnamefont{Bernl{\"o}hr}},
  \bibinfo{author}{\bibfnamefont{C.}~\bibnamefont{Boisson}},
  \bibnamefont{et~al.}, \bibinfo{journal}{Astron. Astrophys.}
  \textbf{\bibinfo{volume}{435}}, \bibinfo{pages}{L17}
  (\bibinfo{year}{2005}{\natexlab{a}}).

\bibitem[{\citenamefont{Aharonian
  et~al.}(2009{\natexlab{a}})\citenamefont{Aharonian, Akhperjanian, Anton,
  Almeida, Bazer-Bachi, Becherini, Behera, Bernl{\"o}hr, Boisson, Bochow
  et~al.}}]{Aharonian:2009p729}
\bibinfo{author}{\bibfnamefont{F.}~\bibnamefont{Aharonian}},
  \bibinfo{author}{\bibfnamefont{A.~G.} \bibnamefont{Akhperjanian}},
  \bibinfo{author}{\bibfnamefont{G.}~\bibnamefont{Anton}},
  \bibinfo{author}{\bibfnamefont{U.~B.~D.} \bibnamefont{Almeida}},
  \bibinfo{author}{\bibfnamefont{A.~R.} \bibnamefont{Bazer-Bachi}},
  \bibinfo{author}{\bibfnamefont{Y.}~\bibnamefont{Becherini}},
  \bibinfo{author}{\bibfnamefont{B.}~\bibnamefont{Behera}},
  \bibinfo{author}{\bibfnamefont{K.}~\bibnamefont{Bernl{\"o}hr}},
  \bibinfo{author}{\bibfnamefont{C.}~\bibnamefont{Boisson}},
  \bibinfo{author}{\bibfnamefont{A.}~\bibnamefont{Bochow}},
  \bibnamefont{et~al.} (\bibinfo{year}{2009}{\natexlab{a}}),
  \eprint{ar$\chi$iv:0906.1247}.

\bibitem[{\citenamefont{Aharonian
  et~al.}(2007{\natexlab{b}})\citenamefont{Aharonian, Akhperjanian,
  Bazer-Bachi, Behera, Beilicke, Benbow, Berge, Bernl{\"o}hr, Boisson, Bolz
  et~al.}}]{Aharonian:2007p712}
\bibinfo{author}{\bibfnamefont{F.}~\bibnamefont{Aharonian}},
  \bibinfo{author}{\bibfnamefont{A.~G.} \bibnamefont{Akhperjanian}},
  \bibinfo{author}{\bibfnamefont{A.~R.} \bibnamefont{Bazer-Bachi}},
  \bibinfo{author}{\bibfnamefont{B.}~\bibnamefont{Behera}},
  \bibinfo{author}{\bibfnamefont{M.}~\bibnamefont{Beilicke}},
  \bibinfo{author}{\bibfnamefont{W.}~\bibnamefont{Benbow}},
  \bibinfo{author}{\bibfnamefont{D.}~\bibnamefont{Berge}},
  \bibinfo{author}{\bibfnamefont{K.}~\bibnamefont{Bernl{\"o}hr}},
  \bibinfo{author}{\bibfnamefont{C.}~\bibnamefont{Boisson}},
  \bibinfo{author}{\bibfnamefont{O.}~\bibnamefont{Bolz}}, \bibnamefont{et~al.},
  \bibinfo{journal}{Astron. Astrophys.} \textbf{\bibinfo{volume}{472}},
  \bibinfo{pages}{489} (\bibinfo{year}{2007}{\natexlab{b}}).

\bibitem[{\citenamefont{Hoppe et~al.}(2009)\citenamefont{Hoppe,
  de~O{\~n}a~Wilhemi, Kh{\'e}lifi, Chaves, de~Jager, Stegmann, and
  Terrier}}]{Hoppe:2009p706}
\bibinfo{author}{\bibfnamefont{S.}~\bibnamefont{Hoppe}},
  \bibinfo{author}{\bibfnamefont{E.}~\bibnamefont{de~O{\~n}a~Wilhemi}},
  \bibinfo{author}{\bibfnamefont{B.}~\bibnamefont{Kh{\'e}lifi}},
  \bibinfo{author}{\bibfnamefont{R.~C.~G.} \bibnamefont{Chaves}},
  \bibinfo{author}{\bibfnamefont{O.~C.} \bibnamefont{de~Jager}},
  \bibinfo{author}{\bibfnamefont{C.}~\bibnamefont{Stegmann}}, \bibnamefont{and}
  \bibinfo{author}{\bibfnamefont{R.}~\bibnamefont{Terrier}}
  (\bibinfo{year}{2009}), \eprint{ar$\chi$iv:0906.5574}.

\bibitem[{\citenamefont{Aharonian
  et~al.}(2006{\natexlab{c}})\citenamefont{Aharonian, Akhperjanian,
  Bazer-Bachi, Beilicke, Benbow, Berge, Bernl{\"o}hr, Boisson, Bolz, Borrel
  et~al.}}]{Aharonian:2006p707}
\bibinfo{author}{\bibfnamefont{F.}~\bibnamefont{Aharonian}},
  \bibinfo{author}{\bibfnamefont{A.~G.} \bibnamefont{Akhperjanian}},
  \bibinfo{author}{\bibfnamefont{A.~R.} \bibnamefont{Bazer-Bachi}},
  \bibinfo{author}{\bibfnamefont{M.}~\bibnamefont{Beilicke}},
  \bibinfo{author}{\bibfnamefont{W.}~\bibnamefont{Benbow}},
  \bibinfo{author}{\bibfnamefont{D.}~\bibnamefont{Berge}},
  \bibinfo{author}{\bibfnamefont{K.}~\bibnamefont{Bernl{\"o}hr}},
  \bibinfo{author}{\bibfnamefont{C.}~\bibnamefont{Boisson}},
  \bibinfo{author}{\bibfnamefont{O.}~\bibnamefont{Bolz}},
  \bibinfo{author}{\bibfnamefont{V.}~\bibnamefont{Borrel}},
  \bibnamefont{et~al.}, \bibinfo{journal}{Astron. Astrophys.}
  \textbf{\bibinfo{volume}{460}}, \bibinfo{pages}{743}
  (\bibinfo{year}{2006}{\natexlab{c}}).

\bibitem[{\citenamefont{Aharonian
  et~al.}(2009{\natexlab{b}})\citenamefont{Aharonian, Akhperjanian, Almeida,
  Bazer-Bachi, Behera, Beilicke, Benbow, Bernl{\"o}hr, Boisson, Bochow
  et~al.}}]{Aharonian:2009p709}
\bibinfo{author}{\bibfnamefont{F.}~\bibnamefont{Aharonian}},
  \bibinfo{author}{\bibfnamefont{A.}~\bibnamefont{Akhperjanian}},
  \bibinfo{author}{\bibfnamefont{U.~D.} \bibnamefont{Almeida}},
  \bibinfo{author}{\bibfnamefont{A.}~\bibnamefont{Bazer-Bachi}},
  \bibinfo{author}{\bibfnamefont{B.}~\bibnamefont{Behera}},
  \bibinfo{author}{\bibfnamefont{M.}~\bibnamefont{Beilicke}},
  \bibinfo{author}{\bibfnamefont{W.}~\bibnamefont{Benbow}},
  \bibinfo{author}{\bibfnamefont{K.}~\bibnamefont{Bernl{\"o}hr}},
  \bibinfo{author}{\bibfnamefont{C.}~\bibnamefont{Boisson}},
  \bibinfo{author}{\bibfnamefont{A.}~\bibnamefont{Bochow}},
  \bibnamefont{et~al.}, \bibinfo{journal}{Astrophys. J.}
  \textbf{\bibinfo{volume}{692}}, \bibinfo{pages}{1500}
  (\bibinfo{year}{2009}{\natexlab{b}}).

\bibitem[{\citenamefont{Aharonian
  et~al.}(2006{\natexlab{d}})\citenamefont{Aharonian, Akhperjanian,
  Bazer-Bachi, Beilicke, Benbow, Berge, Bernl{\"o}hr, Boisson, Bolz, Borrel
  et~al.}}]{Aharonian:2006p738}
\bibinfo{author}{\bibfnamefont{F.}~\bibnamefont{Aharonian}},
  \bibinfo{author}{\bibfnamefont{A.~G.} \bibnamefont{Akhperjanian}},
  \bibinfo{author}{\bibfnamefont{A.~R.} \bibnamefont{Bazer-Bachi}},
  \bibinfo{author}{\bibfnamefont{M.}~\bibnamefont{Beilicke}},
  \bibinfo{author}{\bibfnamefont{W.}~\bibnamefont{Benbow}},
  \bibinfo{author}{\bibfnamefont{D.}~\bibnamefont{Berge}},
  \bibinfo{author}{\bibfnamefont{K.}~\bibnamefont{Bernl{\"o}hr}},
  \bibinfo{author}{\bibfnamefont{C.}~\bibnamefont{Boisson}},
  \bibinfo{author}{\bibfnamefont{O.}~\bibnamefont{Bolz}},
  \bibinfo{author}{\bibfnamefont{V.}~\bibnamefont{Borrel}},
  \bibnamefont{et~al.}, \bibinfo{journal}{Science}
  \textbf{\bibinfo{volume}{314}}, \bibinfo{pages}{1424}
  (\bibinfo{year}{2006}{\natexlab{d}}).

\bibitem[{\citenamefont{Aharonian
  et~al.}(2007{\natexlab{c}})\citenamefont{Aharonian, Akhperjanian,
  Bazer-Bachi, Beilicke, Benbow, Berge, Bernl{\"o}hr, Boisson, Bolz, Borrel
  et~al.}}]{Aharonian:2007p728}
\bibinfo{author}{\bibfnamefont{F.}~\bibnamefont{Aharonian}},
  \bibinfo{author}{\bibfnamefont{A.~G.} \bibnamefont{Akhperjanian}},
  \bibinfo{author}{\bibfnamefont{A.~R.} \bibnamefont{Bazer-Bachi}},
  \bibinfo{author}{\bibfnamefont{M.}~\bibnamefont{Beilicke}},
  \bibinfo{author}{\bibfnamefont{W.}~\bibnamefont{Benbow}},
  \bibinfo{author}{\bibfnamefont{D.}~\bibnamefont{Berge}},
  \bibinfo{author}{\bibfnamefont{K.}~\bibnamefont{Bernl{\"o}hr}},
  \bibinfo{author}{\bibfnamefont{C.}~\bibnamefont{Boisson}},
  \bibinfo{author}{\bibfnamefont{O.}~\bibnamefont{Bolz}},
  \bibinfo{author}{\bibfnamefont{V.}~\bibnamefont{Borrel}},
  \bibnamefont{et~al.}, \bibinfo{journal}{Astron. Astrophys.}
  \textbf{\bibinfo{volume}{467}}, \bibinfo{pages}{1075}
  (\bibinfo{year}{2007}{\natexlab{c}}).

\bibitem[{\citenamefont{Aharonian
  et~al.}(2006{\natexlab{e}})\citenamefont{Aharonian, Akhperjanian,
  Bazer-Bachi, Beilicke, Benbow, Berge, Bernl{\"o}hr, Boisson, Bolz, Borrel
  et~al.}}]{Aharonian:2006p725}
\bibinfo{author}{\bibfnamefont{F.}~\bibnamefont{Aharonian}},
  \bibinfo{author}{\bibfnamefont{A.~G.} \bibnamefont{Akhperjanian}},
  \bibinfo{author}{\bibfnamefont{A.~R.} \bibnamefont{Bazer-Bachi}},
  \bibinfo{author}{\bibfnamefont{M.}~\bibnamefont{Beilicke}},
  \bibinfo{author}{\bibfnamefont{W.}~\bibnamefont{Benbow}},
  \bibinfo{author}{\bibfnamefont{D.}~\bibnamefont{Berge}},
  \bibinfo{author}{\bibfnamefont{K.}~\bibnamefont{Bernl{\"o}hr}},
  \bibinfo{author}{\bibfnamefont{C.}~\bibnamefont{Boisson}},
  \bibinfo{author}{\bibfnamefont{O.}~\bibnamefont{Bolz}},
  \bibinfo{author}{\bibfnamefont{V.}~\bibnamefont{Borrel}},
  \bibnamefont{et~al.}, \bibinfo{journal}{Astron. Astrophys.}
  \textbf{\bibinfo{volume}{456}}, \bibinfo{pages}{245}
  (\bibinfo{year}{2006}{\natexlab{e}}).

\bibitem[{\citenamefont{Aharonian
  et~al.}(2005{\natexlab{b}})\citenamefont{Aharonian, Akhperjanian,
  Bazer-Bachi, Beilicke, Benbow, Berge, Bernl{\"o}hr, Boisson, Bolz, Borrel
  et~al.}}]{Aharonian:2005p643}
\bibinfo{author}{\bibfnamefont{F.}~\bibnamefont{Aharonian}},
  \bibinfo{author}{\bibfnamefont{A.~G.} \bibnamefont{Akhperjanian}},
  \bibinfo{author}{\bibfnamefont{A.~R.} \bibnamefont{Bazer-Bachi}},
  \bibinfo{author}{\bibfnamefont{M.}~\bibnamefont{Beilicke}},
  \bibinfo{author}{\bibfnamefont{W.}~\bibnamefont{Benbow}},
  \bibinfo{author}{\bibfnamefont{D.}~\bibnamefont{Berge}},
  \bibinfo{author}{\bibfnamefont{K.}~\bibnamefont{Bernl{\"o}hr}},
  \bibinfo{author}{\bibfnamefont{C.}~\bibnamefont{Boisson}},
  \bibinfo{author}{\bibfnamefont{O.}~\bibnamefont{Bolz}},
  \bibinfo{author}{\bibfnamefont{V.}~\bibnamefont{Borrel}},
  \bibnamefont{et~al.}, \bibinfo{journal}{Astron. Astrophys.}
  \textbf{\bibinfo{volume}{437}}, \bibinfo{pages}{L7}
  (\bibinfo{year}{2005}{\natexlab{b}}).

\bibitem[{\citenamefont{Albert et~al.}(2007)\citenamefont{Albert, Aliu,
  Anderhub, Antoranz, Armada, Baixeras, Barrio, Bartko, Bastieri, Becker
  et~al.}}]{Albert:2007p710}
\bibinfo{author}{\bibfnamefont{J.}~\bibnamefont{Albert}},
  \bibinfo{author}{\bibfnamefont{E.}~\bibnamefont{Aliu}},
  \bibinfo{author}{\bibfnamefont{H.}~\bibnamefont{Anderhub}},
  \bibinfo{author}{\bibfnamefont{P.}~\bibnamefont{Antoranz}},
  \bibinfo{author}{\bibfnamefont{A.}~\bibnamefont{Armada}},
  \bibinfo{author}{\bibfnamefont{C.}~\bibnamefont{Baixeras}},
  \bibinfo{author}{\bibfnamefont{J.~A.} \bibnamefont{Barrio}},
  \bibinfo{author}{\bibfnamefont{H.}~\bibnamefont{Bartko}},
  \bibinfo{author}{\bibfnamefont{D.}~\bibnamefont{Bastieri}},
  \bibinfo{author}{\bibfnamefont{J.~K.} \bibnamefont{Becker}},
  \bibnamefont{et~al.}, \bibinfo{journal}{Astron. Astrophys.}
  \textbf{\bibinfo{volume}{474}}, \bibinfo{pages}{937} (\bibinfo{year}{2007}).

\bibitem[{\citenamefont{Aharonian et~al.}(2008)\citenamefont{Aharonian,
  Akhperjanian, Almeida, Bazer-Bachi, Behera, Beilicke, Benbow, Bernl{\"o}hr,
  Boisson, Borrel et~al.}}]{Aharonian:2008p733}
\bibinfo{author}{\bibfnamefont{F.}~\bibnamefont{Aharonian}},
  \bibinfo{author}{\bibfnamefont{A.~G.} \bibnamefont{Akhperjanian}},
  \bibinfo{author}{\bibfnamefont{U.~B.~D.} \bibnamefont{Almeida}},
  \bibinfo{author}{\bibfnamefont{A.~R.} \bibnamefont{Bazer-Bachi}},
  \bibinfo{author}{\bibfnamefont{B.}~\bibnamefont{Behera}},
  \bibinfo{author}{\bibfnamefont{M.}~\bibnamefont{Beilicke}},
  \bibinfo{author}{\bibfnamefont{W.}~\bibnamefont{Benbow}},
  \bibinfo{author}{\bibfnamefont{K.}~\bibnamefont{Bernl{\"o}hr}},
  \bibinfo{author}{\bibfnamefont{C.}~\bibnamefont{Boisson}},
  \bibinfo{author}{\bibfnamefont{V.}~\bibnamefont{Borrel}},
  \bibnamefont{et~al.}, \bibinfo{journal}{Astron. Astrophys.}
  \textbf{\bibinfo{volume}{490}}, \bibinfo{pages}{685} (\bibinfo{year}{2008}).

\bibitem[{\citenamefont{Aharonian
  et~al.}(2009{\natexlab{c}})\citenamefont{Aharonian, Akhperjanian, Anton,
  de~Almeida, Bazer-Bachi, Becherini, Behera, Benbow, Bernl{\"o}hr, Boisson
  et~al.}}]{Aharonian:2009p734}
\bibinfo{author}{\bibfnamefont{F.}~\bibnamefont{Aharonian}},
  \bibinfo{author}{\bibfnamefont{A.~G.} \bibnamefont{Akhperjanian}},
  \bibinfo{author}{\bibfnamefont{G.}~\bibnamefont{Anton}},
  \bibinfo{author}{\bibfnamefont{U.~B.} \bibnamefont{de~Almeida}},
  \bibinfo{author}{\bibfnamefont{A.~R.} \bibnamefont{Bazer-Bachi}},
  \bibinfo{author}{\bibfnamefont{Y.}~\bibnamefont{Becherini}},
  \bibinfo{author}{\bibfnamefont{B.}~\bibnamefont{Behera}},
  \bibinfo{author}{\bibfnamefont{W.}~\bibnamefont{Benbow}},
  \bibinfo{author}{\bibfnamefont{K.}~\bibnamefont{Bernl{\"o}hr}},
  \bibinfo{author}{\bibfnamefont{C.}~\bibnamefont{Boisson}},
  \bibnamefont{et~al.}, \bibinfo{journal}{Astrophys. J. Lett.}
  \textbf{\bibinfo{volume}{695}}, \bibinfo{pages}{L40}
  (\bibinfo{year}{2009}{\natexlab{c}}).

\bibitem[{\citenamefont{Meintjes et~al.}(1992)\citenamefont{Meintjes,
  Raubenheimer, de~Jager, Brink, Nel, North, van Urk, and
  Visser}}]{Meintjes:1992p647}
\bibinfo{author}{\bibfnamefont{P.}~\bibnamefont{Meintjes}},
  \bibinfo{author}{\bibfnamefont{B.}~\bibnamefont{Raubenheimer}},
  \bibinfo{author}{\bibfnamefont{O.}~\bibnamefont{de~Jager}},
  \bibinfo{author}{\bibfnamefont{C.}~\bibnamefont{Brink}},
  \bibinfo{author}{\bibfnamefont{H.~I.} \bibnamefont{Nel}},
  \bibinfo{author}{\bibfnamefont{A.~R.} \bibnamefont{North}},
  \bibinfo{author}{\bibfnamefont{G.}~\bibnamefont{van Urk}}, \bibnamefont{and}
  \bibinfo{author}{\bibfnamefont{B.}~\bibnamefont{Visser}},
  \bibinfo{journal}{Astrophys. J.} \textbf{\bibinfo{volume}{401}},
  \bibinfo{pages}{325} (\bibinfo{year}{1992}).

\bibitem[{\citenamefont{Gould and Schr{\'e}der}(1966)}]{GRH}
\bibinfo{author}{\bibfnamefont{R.~J.} \bibnamefont{Gould}} \bibnamefont{and}
  \bibinfo{author}{\bibfnamefont{G.}~\bibnamefont{Schr{\'e}der}},
  \bibinfo{journal}{Phys. Rev. Lett.} \textbf{\bibinfo{volume}{16}},
  \bibinfo{pages}{252} (\bibinfo{year}{1966}).

\bibitem[{\citenamefont{Colladay and Kosteleck{\'y}}(2001)}]{CKcross}
\bibinfo{author}{\bibfnamefont{D.}~\bibnamefont{Colladay}} \bibnamefont{and}
  \bibinfo{author}{\bibfnamefont{V.~A.} \bibnamefont{Kosteleck{\'y}}},
  \bibinfo{journal}{Phys. Lett. B} \textbf{\bibinfo{volume}{511}},
  \bibinfo{pages}{209} (\bibinfo{year}{2001}).

\end{thebibliography}

\end{document}